\documentstyle[preprint,aps]{revtex}
\tighten
\begin{document}
\preprint{Preprint Number:
\parbox[t]{50mm}{ADP-94-24/T164 \\
		 hep-ph/9501262}  }
\draft
\title{Solving the Bethe-Salpeter Equation for Scalar \\
       Theories in Minkowski Space}
\author{
  Kensuke Kusaka\footnote{e-mail: {\it kkusaka@physics.adelaide.edu.au}}
	  and
  Anthony G. Williams\footnote{e-mail: {\it awilliam@physics.adelaide.edu.au}}
  \vspace*{2mm} }

\address{
    Department of Physics and Mathematical Physics, \\
    University of Adelaide, \\
    South Australia 5005, Australia
}
%
\maketitle

\begin{abstract}
The Bethe-Salpeter (BS) equation for scalar-scalar bound
states in scalar theories without derivative coupling
is formulated and solved in Minkowski space.
This is achieved using the perturbation theory integral representation
(PTIR), which allows these amplitudes to be
expressed as integrals over weight functions and known singularity
structures and hence allows us to convert the BS equation
into an integral equation involving weight functions.
We obtain numerical solutions using this formalism
for a number of scattering kernels to illustrate the generality of the
approach.  It applies even when the na\"{\i}ve
Wick rotation is invalid. As a check we verify, for example, that
this method applied to the special case of the massive ladder exchange
kernel reproduces the same results as are obtained by Wick rotation.
\end{abstract}


\section{Introduction}

There has been considerable recent interest in
covariant descriptions
of bound states, for example, in conjunction with model calculations
of high-energy processes such as deep inelastic scattering.
A fully covariant description of composite
bound states is essential for the understanding of hadronic structure
over the full range of available momentum transfer.
In a relativistic field theory
the two-body component of a bound state is described
by the appropriate proper (i.e., one-particle irreducible)
three-point vertex function or, equivalently, by the Bethe-Salpeter (BS)
amplitude.  An extensive review of the BS equation has been given by
Nakanishi \cite{Nakanishi_survey}.  Other reviews of BS equation
studies and many further references can be be found in
Refs.~\cite{Nakanishi_WC,Seto,Murota,BSE_refs}.
The BS amplitude satisfies an integral equation
whose kernel has singularities due to the Minkowski metric.
The resultant solutions are not functions but mathematical distributions.
The singularity structure of these distributions
makes it difficult to handle the BS equation
numerically in Minkowski space.

We illustrate the BS equation for a scalar
theory in Fig.~1, where $\Phi(p,P)$
is the BS amplitude, $P\equiv p_1 + p_2$ is the total four-momentum of
the bound state
and $p\equiv \eta_2 p_1 -\eta_1 p_2$ is the relative four-momentum for
the two scalar constituents.  We have then
$M=\sqrt{P^2}$ for the bound state mass and also
$\eta_1+\eta_2=1$, but otherwise
the choice of the two positive real numbers $\eta_1$ and
$\eta_2$ is arbitrary.
Note that $p_1=\eta_1 P+p$ and $p_2=\eta_2 P-p$.  In the
nonrelativistic limit the natural choice is $\eta_{1,2}=m_{1,2}/(m_1+m_2)$,
e.g., $\eta_1=\eta_2=1/2$ for the equal mass case.  This is the choice that
we will make here.

The renormalized constituent scalar propagators are
$D(p_{1,2}^2)$ and
$K(p,q;P)$ is the renormalized scattering kernel.  For example, in simple
ladder approximation in a $\phi^2\sigma$ model we would have
$K(p,q;P)=(ig)(iD_\sigma([p-q]^2))(ig)$, where
$D_\sigma(p^2)=1/(p^2-m_\sigma^2+i\epsilon)$ and $m_\sigma$ is the
$\sigma$-particle mass.  Note that the corresponding
proper (i.e, one-particle irreducible) vertex for the bound state
is related to the BS amplitude by $\Phi=(iD)(i\Gamma)(iD)$.  We
follow standard conventions in our definitions of quantities,
(see Ref.~\cite{B+D,I+Z} and also, e.g., \cite{TheReview}).  Thus the
BS equation
for any scalar theory can be written as
\begin{equation}
i\Gamma(p_1,p_2) = \int{d^4q\over(2\pi)^4}
        (iD(q_1^2))~
        (i\Gamma(q_1,q_2))~(iD(q_2^2))~K(p,q;P)\, ,
	\label{BSE_Gamma}
\end{equation}
where similarly to $p_1$ and $p_2$ we have defined
$q_1=\eta_1 P+q$ and $q_2=\eta_2 P-q$.
Equivalently in terms of the BS amplitude we can write
\begin{eqnarray}
  D(p_1^2)^{-1}~\Phi(p,P)~D(p_2^2)^{-1}
        &=&-\int{d^4q\over(2\pi)^4}~\Phi(q,P)~K(p,q;P)
	\label{BSE_K} \\
        &\equiv&~\int{d^4q\over(2\pi)^4 i}~\Phi(q,P)~I(p,q;P)\, .
	\label{BSE}
\end{eqnarray}
where the kernel function defined by $I(p,q;P)\equiv-iK(p,q;P)$ is the form
typically used by Nakanishi \cite{Nakanishi_survey}.  In ladder approximation
for a $\phi^2\sigma$ model we see for example that
$I(p,q;P)=g^2/(m_\sigma^2-p^2-i\epsilon)$.  For ease of comparison with earlier
work we will use the latter form of the BS equation here, i.e.,
Eq.~(\ref{BSE}).

One approach to dealing with the difficulties presented by the Minkowski-space
metric is to perform an analytic continuation in the
relative-energy variable $p^0$, which is the so-called ``Wick rotation''
\cite{Wick}.  This has been widely used as a means of solving model
BS equations, e.g., see Ref.~\cite{TheReview} and references therein.
The special case of the ladder BS equation is solved as a function
of Euclidean
relative momentum in the standard treatment.  This is possible since
in the ladder approximation the kernel is independent of the
the total four-momentum $P^\mu$.
The difficulties associated with this approach in the general case
arise from the fact that
since the total four-momentum must remain timelike then
$P\cdot p$ becomes complex.
In addition, when one uses a ``dressed'' propagator
for the constituent particles or more complicated kernels
in the BS equation, the validity of the Wick rotation becomes
highly  nontrivial.  For example, essentially all of
the dressed propagators studied previously in Wick-rotated
Dyson-Schwinger equation
approach contain complex ``ghost'' poles\cite{TheReview}.
Hence, the na\"{\i}ve Wick rotation obtained by the simple
transcription of Minkowski metric to the Euclidean metric
and vice versa is not valid in general.
So while the Euclidean-based approaches certainly play a very important role
and can be useful in model calculations, it is
preferable to formulate and solve the BS equation directly in
Minkowski space.
Here we present a method to solve the BS equation without Wick
rotation by making use of the perturbation theory integral representation
(PTIR) for the BS amplitude\cite{Nakanishi_graph}.

The PTIR method was first used in conjunction with a Wick rotation
by Wick and Cutkosky for a scalar-scalar bound state
with a massless scalar exchange
in the ladder approximation\cite{Wick,Cutkosky}.
This is now commonly referred to as the Wick-Cutkosky model.
They solved the BS equation in terms
of a single variable integral representation,
which is a special case of the PTIR.
Wanders \cite{Wanders} first introduced the more general
two-variable integral representation
to solve the BS equation for a scalar-scalar bound state
with a massive scalar exchange in the ladder approximation.
It was shown that all invariant scalar-scalar BS amplitudes
have this integral representation
in the ladder approximation\cite{I+M,Nakanishi63}
and the weight function for the corresponding BS amplitude was
solved formally by means of Fredholm theory\cite{Nakanishi63,Sato}.
This two-variable integral representation was independently proposed
for the vertex function on the basis of axiomatic field theory
\cite{Fainberg,D+G+S,Ida}.
Following this proposal of an integral representation
for the vertex function,
Nakanishi made
a detailed and systematic study of the PTIR\cite{Nakanishi_graph}.

The PTIR is a natural extension of the spectral representation for a two-point
Green's function to
an $n$-point function in a relativistic field theory.  Since
the Feynman parametric integral always exists for any Feynman diagram
in perturbative calculations,
one can always define the integral representation
such that the number of independent integration parameters is equal
to that of invariant squares of external momenta.
Each Feynman diagram contributes to the
weight distribution of the parametric integral, so that
the complete weight function for the renormalized $n$-point function
is identical to the infinite sum of Feynman diagrams for the renormalized
Lagrangian of the theory.  Hence, we see that
the PTIR of a particular renormalized $n$-point function is an integral
representation of the corresponding infinite sum of Feynman
diagrams for the renormalized theory with $n$ fixed external lines.

This paper is organized as follows.  In Sec.~\ref{Representation},
we introduce the PTIR of the renormalized
scattering kernel for scalar models without derivative coupling.
We also introduce the PTIR for the BS amplitude
itself.  We discuss the structure
of the kernel weight function and the BS amplitude.
In Sec.~\ref{Integral} an integral equation for the BS amplitude
weight function
is derived and expressed in terms of a new type of kernel function,
resulting from the kernel weight function
and the constituent scalar propagators.  The structure of this new kernel
function are discussed in Sec.~\ref{Singularity}.  Numerical results are
presented in Sec.~\ref{Results} and we present our conclusions and directions
for future work in Sec.~\ref{Conclusions}.

\section{PTIR for Scalar Theories}
\label{Representation}

In this treatment we will limit ourselves to studies of bound states of scalar
particles interacting through a scalar kernel without derivative coupling.
Let $\phi(x)$ be the field operator for the constituent
scalar particles having a renormalized mass $m$.
We will define $\mu$ such that $m+\mu$ is the threshold for particle
production in the single $\phi$ channel.
For example, for a $\phi^2\sigma$ model where
the renormalized $\sigma$-particle mass satisfies $m_\sigma<2m$ or
$m_\sigma>2m$, then we would have $\mu=m_\sigma$ or $\mu=2m$ respectively.
The renormalized propagator for the $\phi$-particle
can be written in the following spectral form;
\begin{equation}
        D(q)=-\left({1 \over m^2 - q^2 -i\epsilon}
                + \int_{(m+\mu)^2}^\infty d\alpha
                \frac{\rho_\phi(\alpha)}{\alpha-q^2-i\epsilon}\right),
        \label{phi_prop}
\end{equation}
where $\rho_\phi(\alpha)$ is the renormalized spectral function.
Note that $\rho_\phi(\alpha)\ge 0$, (see, e.g.,
\cite{B+D}).

Following the conventions of Ref.~\cite{I+Z}, (e.g., pp.~481-487),
we can also define centre-of-momentum and relative coordinates
$X\equiv \eta_1x_1+\eta_2x_2$ and $x\equiv x_1-x_2$ such that
$x_1=X+\eta_2x$, $x_2=X-\eta_1x$, and
$P\cdot X + p\cdot x = p_1\cdot x_1 + p_2\cdot x_2$.  Hence
the Bethe-Salpeter amplitude $\Phi(p,P)$ for the bound state
of two $\phi$-particles having the total momentum $P\equiv p_1 +p_2$ and
relative momentum $p\equiv (\eta_2p_1-\eta_1p_2)$ can be defined as
\begin{equation}
	\langle 0|T\phi(x_1)\phi(x_2)|P\rangle
	= e^{-iP\cdot X}\langle 0|T\phi(\eta_2 x)\phi(-\eta_1 x)|P\rangle
	= e^{-iP\cdot X}
	\int{d^4p\over (2\pi)^4} e^{-ip\cdot x} \Phi(p,P)\;,
\label{BS_amp}
\end{equation}
where we have made use of the translational invariance of the BS amplitude.
Equivalently, we can write
\begin{equation}
	\Phi(p,P)
	=e^{iP\cdot X}\int d^4x e^{ip\cdot x}
		\langle 0|T\phi(x_1)\phi(x_2)|P\rangle
	=\int d^4x e^{ip\cdot x}
		\langle 0|T\phi(\eta_2 x)\phi(-\eta_1 x)|P\rangle\;.
\label{BS_FT}
\end{equation}
Note that the bound states are normalized such that
$\langle P|P'\rangle=2\omega_P(2\pi)^3\delta^3(\vec P'-\vec P)$,
where $\omega_P\equiv ({\vec P}^2+M^2)^{1/2}$ with $M$ the bound state mass.
For a positive energy bound state we must have $P^2=M^2$,
$0 < P^2 \le (2m)^2$, and $P^0 > 0$.
The normalization condition for the BS amplitude is given by
\begin{equation}
\int \frac{d^4p}{(2\pi)^4}\int \frac{d^4q}{(2\pi)^4}
	\bar\Phi(q,P)\frac{\partial}{\partial P_\mu}
	\left\{D^{-1}(p_1^2)D^{-1}(p_2^2)(2\pi)^4\delta^4(p-q)
	+K(p,q;P)\right\}\Phi(p,P)=2iP^\mu\;,
\label{normalization}
\end{equation}
where the conjugate BS amplitude $\bar\Phi(p,P)$ is defined by
\begin{equation}
	\bar\Phi(p,P)=e^{-iP\cdot X}\int d^4x e^{-ip\cdot x}
		\langle P|T\phi^\dagger(x_1)\phi^\dagger(x_2)|0\rangle
	=\int d^4x e^{-ip\cdot x}
	\langle P|T\phi^\dagger(\eta_2 x)\phi^\dagger(-\eta_1 x)|0\rangle\;.
\end{equation}

\subsection{PTIR for Scattering Kernel}
The scattering kernel $I(p,q;P)\equiv -iK(p,q;P)$ describes the process
$\phi\phi \rightarrow \phi\phi$, where
$p$ and $q$ are the initial and final relative momenta
respectively.  It is given by the infinite series of Feynman diagrams
which are two-particle irreducible
with respect to the initial and final pairs of constituent $\phi$
particles.
For purely scalar theories without derivative coupling we have
the formal expression for the full
renormalized scattering kernel
\begin{eqnarray}
    I(p,q;P) &=&
    \int\limits_{0}^{\infty } d\gamma\int\limits_{\Omega } d\vec\xi
     \left\{ \frac{ \rho_{st} (\gamma,\vec\xi) }
    { \gamma - \left[ \sum_{i=1}^{4}\xi_iq_i^2+\xi_5s+\xi_6t
     \right]-i\epsilon }
    \right.
    + \frac{ \rho_{tu}(\gamma,\vec\xi) }
    { \gamma - \left[ \sum_{i=1}^{4}\xi_iq_i^2+\xi_5t+\xi_6u \right]
     -i\epsilon }
    \nonumber\\
    &  & \left. + \frac{\rho_{us}(\gamma,\vec\xi) }
    {\gamma-\left[\sum_{i=1}^{4}\xi_iq_i^2+\xi_5u+\xi_6s\right]-i\epsilon}
    \right\}\;,
    \label{krnl_PTIR}
\end{eqnarray}
where $q_i^2$ is the 4-momentum squared carried by $\phi_i$
and $s$,$t$ and $u$ are the usual Mandelstam variables.
This expression has been derived by Nakanishi (PTIR)\cite{Nakanishi_graph}.
Since only six of these are independent due to the relation
$q_1^2+q_2^2+q_3^2+q_4^2=s+t+u$, this is also the number of independent
$\xi$-parameters.  Hence, we need only introduce
the ``mass'' parameter $\gamma$ and
six dimensionless Feynman parameters $\xi_i$ with a constraint.
The symbol $\Omega$ denotes the integral region of
$\xi_i$ such that $\Omega\equiv\{\xi_i \,| \,0\leq\xi_i\leq 1, \,
\sum\xi_i=1 \, (i=1,\dots, 6)\}$.  The ``mass'' parameter $\gamma$
represents a spectrum of the scattering kernel.  The function
$\rho_{\rm ch}(\gamma,\vec\xi)$ gives the weight
of the spectrum arising from three different
channels which can be denoted ${\rm ch}= \{st\},\{tu\},\{us\}$.
This PTIR expression follows since any perturbative Feynman diagram for the
scattering kernel
can be written in this form, and hence this must also be true of their
sum.  In a perturbative calculation the weight function
$\rho_{\rm ch}$ is calculable as a power series of the coupling
constant for a given interaction Lagrangian.
For more general theories involving, e.g., fermions and/or derivative
couplings, the numerator of Eq.~(\ref{krnl_PTIR})
will also contain momenta in general.  This would considerably complicate
the procedure to be presented here and the problem of how
to extend this approach to those cases is as yet unresolved.
One can impose additional support properties
on the kernel weight functions $\rho_{\rm ch}(\gamma,\vec\xi)$
by analyzing general Feynman integrals for the theory of interest
\cite{Nakanishi_graph}.  However, we do not impose any further conditions
at this stage since we wish to incorporate more general cases such as
separable kernels, which cannot
be written as combinations of ordinary Feynman diagrams.

To illustrate the approach we will discuss two explicit
examples here: \\
	(a) Scalar-scalar ladder model with massive scalar exchange:
	The simple $t$-channel one-$\sigma$-exchange kernel is given by
	\begin{equation}
		I(p,q;P)=\frac{g^2}{m_\sigma^2-(p-q)^2-i\epsilon}
		\label{pure_ladder_kernel}
	\end{equation}
	The BS equation with this kernel together with perturbative
	 propagator $D^0$ is often referred to as the
	``scalar-scalar ladder model''\cite{Nakanishi_survey}.
	The Wick rotated BS equation for this kernel has been studied
	numerically\cite{L+M}.  We use this kernel as a check of our
	calculations.\\
	(b) Generalized ladder kernel:
	A sum of the one-$\sigma$-exchange kernel
	Eq.~(\ref{pure_ladder_kernel})
	and a generalized kernel with fixed kernel parameter sets
	$\{\gamma^{(i)}, \vec\xi^{(i)}\}$.
	After the Wick rotation this kernel
	becomes complex due to the $p\cdot P$ and $q\cdot P$ terms, so that
	solving the BS amplitude as a function of Euclidean relative momentum
	would be very difficult in this case.

\subsection{PTIR for BS Amplitude}
\label{sec_PTIR_BS}

Let us now consider the PTIR of the BS amplitude itself.
Since the BS amplitude in the center-of-momentum
rest frame [i.e., with $P^\mu=(M,\vec 0)$] forms
an irreducible representation of the $O(3)$ rotation
group, we can label all of the
bound states with the usual 3-dimensional angular momentum quantum
numbers $\ell$ and $\ell_z$.
We can thus construct the integral representation
of the partial wave BS amplitude in the rest frame and the PTIR
automatically allows us to then boost to an arbitrary frame.

For simplicity, let us begin by considering the $s$-wave amplitude.
In order to apply the PTIR to the BS amplitude, it is convenient
to first consider the equivalent vertex function $\Gamma(p,P)$
defined such that
$i\Phi(p,P)=D((P/2)+p)\Gamma(p,P)D((P/2)-p)$.
The vertex function $\Gamma(p,P)$ is a one-particle-irreducible
3-point Green's function, which can be expressed as an infinite
sum of Feynman diagrams and hence in terms of PTIR has the form
derived by Nakanishi \cite{Nakanishi_graph}
\begin{equation}
	\Gamma(p,P)=\int\limits_{0}^{\infty}d\kappa
	\prod_{i=1}^3\int\limits_{0}^{1}dz_i\delta(1-\sum\limits_{i=1}^3z_i)
	\frac{\rho_3(\kappa,\vec z)}
	{\kappa-\left(z_1(\frac{P}{2}+p)^2 +
	z_2 (\frac{P}{2}-p)^2 + z_3 P^2\right)
	-i\epsilon}\;.
	\label{vtexPTIR1}
\end{equation}
In contrast to the case for the scattering kernel,
all invariant squares of momenta are independent and
so a single weight function $\rho_3(\kappa,\vec z)$ is sufficient to
describe the sum of all allowed
Feynman diagrams.  From the perturbative analysis of
Feynman graphs for the vertex function one obtains
the following support property for the vertex weight function
\begin{eqnarray}
	\rho_3(\kappa,\vec z)=0, &
		\quad\text{unless}\quad  \kappa\ge\text{max}[
	&(z_1+z_2)(m+\mu)^2,
	\nonumber \\
	 &  & 	z_1(m+\mu)^2+z_2(m-\mu)^2+z_3(2m)^2,
	\nonumber \\
	 &  & z_1(m-\mu)^2+z_2(m+\mu)^2+z_3(2m)^2 ]\;,
	\label{Supp_vtexPTIR1}
\end{eqnarray}
where ``max'' is taken separately for each fixed set of $\vec z$.
For the vertex function with a timelike total momentum satisfying
$0 < P^2 < 4m^2$, it follows that Eq.~(\ref{vtexPTIR1}) reduces to
the following two-variable representation
\begin{equation}
	\Gamma(p,P)=\int\limits_{0}^{\infty}d\beta
	\int\limits_{-1}^{1}d\zeta\frac{\tilde\rho_3(\beta, \zeta)}
	{\beta-(p + \zeta {P\over 2})^2 -i\epsilon}\;,
	\label{vtexPTIR2}
\end{equation}
with the support property
\begin{equation}
	\tilde\rho_3(\beta, z)=0,\quad\text{unless}\quad
	\beta\ge \left(m+\mu-{1\over 2}(1-|\zeta|)\sqrt{P^2}\right)^2\;.
	\label{Supp_vtexPTIR2}
\end{equation}
This two-variable representation was first proposed
in axiomatic field theory \cite{D+G+S,Ida}.

For simplicity, we will now assume that the constituent
$\phi$-propagators have their free form,
$D(p^2)\to D^0(p^2)=1/(p^2-m^2+i\epsilon)$, or equivalently
$\rho_\phi(\alpha)=0$ in Eq.~(\ref{phi_prop}).  The arguments
can be extended in a straightforward way to the case
$\rho_\phi(\alpha)\neq 0$, but we will not pursue this possibility here.
Then attaching two $\phi$-propagators as external lines
yields the PTIR for the BS amplitude
\begin{equation}
	\Phi(p,P)=-i\int\limits_{-\infty}^{\infty}d\alpha
		\int\limits_{-1}^{1}dz
		\frac{\varphi_n(\alpha,z)}
		{\left[m^2+\alpha-\left(p^2 + z p\cdot P+\frac{P^2}{4}\right)
			-i\epsilon\right]^{n+2}}\;,
	\label{BS_PTIR}
\end{equation}
where we have introduced the non-negative integer dummy parameter $n$.
In order for the expression to be meaningful, we see that we must
have the boundary conditions
\begin{eqnarray}
	& &\lim_{\alpha\rightarrow\infty}
			\frac{\varphi_n(\alpha,z)}{\alpha^{n+1}}=0,\nonumber\\
	& &\varphi_n(\alpha=-\infty,z)=0\;.
  \label{bc_PTIR}
\end{eqnarray}
We see that a partial integration of Eq.~(\ref{BS_PTIR}) with respect to
$\alpha$ connects weight functions with different $n$, i.e.,
\begin{equation}
	\varphi_{n+1}(\alpha,z)=(n+2)\int\limits_{-\infty}^\alpha
					d\alpha'\varphi_{n}(\alpha',z)\;,
	\label{rln_wghtfnc}
\end{equation}
and hence, as stated above, the integer $n$ is an artificial or
dummy parameter.  While the non-negative integer $n$ is arbitrary,
a judicious choice can be advantageous
in numerical calculations, since the larger we take $n$,
the smoother the weight function becomes for a given $\Phi(p,P)$.
{}From the support property of $\tilde\rho_3(\beta, z)$ in
Eq.~(\ref{Supp_vtexPTIR2}), it follows that the weight
function $\varphi_n(\alpha,z)=0$ when
\begin{equation}
	\alpha < \text{min }
		\left[ 0,
			\left(m+\mu-{\sqrt{P^2}\over 2}\right)^2-m^2+
                        {P^2\over 4}
		\right].
	\label{Supp_BS_PTIR}
\end{equation}
This support property can be understood from the
relation between $\tilde\rho_3(\beta, z)$ and $\varphi_n(\alpha,z)$, i.e.,
\begin{eqnarray}
\varphi_n(\alpha,z)&=&\int\limits_{0}^{\infty}d\beta
	\int\limits_{-1}^{1}d\zeta
	\frac{\tilde\rho_3(\beta, \zeta) }
		{\left|\beta-m^2+(1-\zeta^2){P^2\over 4}\right|}
\nonumber\\
	& &\quad \times\theta\left(\frac{\alpha}{\beta-m^2+(1-\zeta^2)
        {P^2\over 4}}\right)
	\theta\left(R(z,\zeta)-\frac{\alpha}{\beta-m^2+(1-\zeta^2)
        {P^2\over 4}}\right),
	\label{rln}
\end{eqnarray}
where $R(\bar z,z)\equiv [(1-\bar z)/(1- z)]\theta(\bar z - z) +
	                [(1+\bar z)/(1+ z)]\theta(z - \bar z )$.
Thus the region of support for $\varphi_n(\alpha,z)$,
which is specified by the step-functions in Eq.~(\ref{rln}), is due to
the singularities of the $\phi$-propagators attached as external lines
to the proper bound-state vertex $\Gamma$.
Note that the integral representation in Eq.~(\ref{BS_PTIR})
and the support
property Eq.~(\ref{Supp_BS_PTIR}) are valid even if we keep the full
nonperturbative $\phi$-propagator $D(p^2)$ by including the
spectral function $\rho_\phi(\alpha)$.
In this case Eq.~(\ref{rln}) for the BS amplitude weight function
$\varphi_n(\alpha,z)$ must be suitably generalized and will
include an integration over each of the spectral
functions from the two $\phi$-propagators.

Now let us consider the BS amplitudes for higher partial waves, i.e,
those for bound states with non-zero angular momentum ($\ell>0$).
To define these, we consider the rest frame of the bound state,
i.e., $P^\mu=(M,\vec 0)$.
The momentum-dependent structures, i.e., the denominators, of the
PTIR's for the bound state proper vertex $\Gamma$ and the BS amplitude
$\Phi$ are independent of
transformations under the little group belonging to the 3-momentum
$\vec p$ in this frame. Hence, for higher partial waves
with angular momentum quantum number
$\ell$ and third component $\ell_z$,
$\Gamma$ and $\Phi$ can be written as the product
of the $\ell$-th order solid harmonic
${\cal Y}_\ell^{\ell_z}(\vec p)=|\vec p|^\ell Y^{\ell_z}_\ell(\vec p)$
and the corresponding PTIR for the scalar (i.e., s-wave) bound state
in this frame \cite{Nakanishi_survey}.
A simple way to understand this is that since the system consists
only of scalar particles and since in this frame there is only one available
3-vector (i.e., $\vec p$), then higher partial wave amplitudes must be
proportional to $Y^{\ell_z}_\ell(\vec p)$.  The solid harmonics are
polynomials of their three arguments and since they have
the self-reproducing property Eq.~(\ref{solid_harm_prop}) \cite{Nakanishi63},
we know that in fact the
amplitudes must be proportional to ${\cal Y}_\ell^{\ell_z}(\vec p)$.
Thus we have in the rest frame for a bound state with
angular momentum quantum numbers $\ell$ and $\ell_z$
\begin{equation}
	\Phi^{[\ell,\ell_z]}(p,P)=
	-i{\cal Y}_\ell^{\ell_z}(\vec p)
		\int\limits_{\alpha_{\rm th}}^{\infty}d\alpha
		\int\limits_{-1}^{1}dz
		\frac{\varphi^{[\ell]}_n(\alpha,z)}
		{\left[m^2+\alpha-\left(p^2 + z p\cdot P +
                \frac{P^2}{4}\right)-i\epsilon\right]^{n+2}},
	\label{ptlw_BS_PTIR}
\end{equation}
where the total momentum $P$ is $P=(\sqrt{P^2}, \vec 0)$ and where
we have introduced the lower bound $\alpha_{\rm th}$
for the integral range of $\alpha$, corresponding
to the minimal value of the right hand side of Eq.~(\ref{Supp_BS_PTIR}).
The boundary conditions for $\varphi^{[\ell]}_n(\alpha,z)$ are identical
to those for the $s$-wave Eq.~(\ref{bc_PTIR}).
The dummy parameter $n$ can always be taken such that
the loop-momentum integral of the BS equation Eq.~(\ref{BSE}) converges.

By performing a straightforward Lorentz boost we have the integral
representation of the partial wave BS amplitude in an arbitrary frame
\begin{equation}
	\Phi^{[\ell,\ell_z]}(p,P)=-i{\cal Y}_\ell^{\ell_z}(\vec{p'})
		\int\limits_{\alpha_{\rm th}}^{\infty}d\alpha
		\int\limits_{-1}^{1}dz
		\frac{\varphi^{[\ell]}_n(\alpha,z)}
		{\left[m^2+\alpha-\left(p^2 + z p\cdot P + P^2/4\right)
			-i\epsilon\right]^{n+2}}\;,
	\label{ptlw_BS_PTIR2}
\end{equation}
where $P$ is an arbitrary timelike 4-vector with $P^2=M^2$
and $p'=\Lambda^{-1}(P)p$.  The Lorentz transformation
$\Lambda(P)$ connects $P$ and the bound-state rest frame 4-vector
$P'=(M,\vec 0)$, i.e, $P = \Lambda(P) P'$.
In the following sections
we will study the BS equation Eq.~(\ref{BSE}) in an arbitrary frame
in terms of this integral representation.

\section{BS Equation for the Weight Function}
\label{Integral}

In this section we will reformulate the BS equation Eq.~(\ref{BSE})
as an integral equation in terms of the weight functions.  This is the
central result of this paper.
Substituting the integral representation of the partial wave BS amplitude
Eq.~(\ref{ptlw_BS_PTIR}) together with the PTIR for the scattering kernel
Eq.~(\ref{krnl_PTIR}) into the BS equation Eq.~(\ref{BSE}),
the right hand side of the BS equation becomes
\begin{eqnarray}
	& &\int{d^4q\over (2\pi)^4 i}
        I(p,q;P)\Phi^{[\ell,\ell_z]}(q,P)\nonumber\\
	& =  & 	-i\;\sum_{\rm ch}\int\limits_{0}^{\infty }
                d\gamma\int\limits_{\Omega}
		d\vec\xi \,\int\limits_{-\infty}^{\infty}
		d\alpha\int\limits_{-1}^{1}dz \,
		\rho_{\rm ch}(\gamma,\vec\xi) \varphi^{[\ell]}_n(\alpha,z)
	\nonumber\\
	& & \times\int{d^4q\over (2\pi)^4i}\,
        	\frac{1}
		{\gamma-\left(
			a_{\rm ch}\,q^2+b_{\rm ch}\,
                                p\cdot q+c_{\rm ch}\,p^2+d_{\rm ch}\,P^2
				+e_{\rm ch}\,q\cdot P+f_{\rm ch}\,p\cdot P
			\right)-i\epsilon
		}\nonumber\\
	& & \qquad\qquad\qquad\times
		\frac{  {\cal Y}_\ell^{\ell_z}(\Lambda^{-1}(P) q)}
		{\left[	m^2+\alpha
			-\left(q^2 + z q\cdot P + P^2/4\right)
			-i\epsilon
		\right]^{n+2}}\;,
	\label{krnl_wght}
\end{eqnarray}
where $\{a_{\rm ch},b_{\rm ch},c_{\rm ch},\dots,f_{\rm ch}\}$ are different
linear combinations of $\xi_i$ in each of the three channels,
{\it ch}$=\{st\},\{tu\},\{us\}$.  They are listed in
Appendix \ref{apdx_a}.  Applying the Feynman parameterization one can
perform the $q$-integral by setting
the dummy parameter $n$ such that $n+1 > l/2$.  This condition ensures
that the $q$-integral is finite.   Due to the self-reproducing
property of the solid harmonics this integral is again proportional
to the solid harmonics ${\cal Y}_\ell^{\ell_z}(\Lambda^{-1}(P) p)$
(see Appendix \ref{apdx_b}).
Now we multiply the propagators $D(P/2+p)D(P/2-p)$ into both sides
and absorb them into the RHS expression using Feynman parameterization.
The BS equation then becomes
\begin{eqnarray}
	\Phi^{[\ell,\ell_z]}(p,P) & = &
	-i{\cal Y}_\ell^{\ell_z}(\Lambda^{-1}(P) p)
	\int\limits_{\alpha_{\rm th}}^{\infty}d
        \bar\alpha\int\limits_{-1}^{1}d\bar z,
	\frac{1}
	{\left[m^2+\bar\alpha
			-\left(p^2 + \bar z p\cdot P + P^2/4\right)
			-i\epsilon\right]^{n+2}
	}
	\nonumber\\
	& & \qquad\times
	\sum_{\rm ch}\int\limits_{0}^{\infty } d\gamma\int\limits_{\Omega}
	\, \rho_{\rm ch}(\gamma,\vec\xi) \,
	\int\limits_{\alpha_{\rm th}}^{\infty}d\alpha\int\limits_{-1}^{1}d
        z \,
		K^{[\ell]}_n(\bar\alpha,\bar z;\alpha,z)
		\,\, \varphi^{[\ell]}_n(\alpha,z)\;.
	\label{krnl_wght2}
\end{eqnarray}
The kernel function $K^{[\ell]}_n(\bar\alpha,\bar z;\alpha,z)$ is defined
with a fixed parameter set of the scattering kernel
$\{a_{\rm ch},b_{\rm ch},c_{\rm ch},\dots,f_{\rm ch}\}$ and
its definition is given in Appendix \ref{apdx_c}.
Comparing Eq.~(\ref{krnl_wght2}) with Eq.~(\ref{ptlw_BS_PTIR2}),
and using the uniqueness theorem of PTIR\cite{Nakanishi_graph},
we obtain the following integral equation for $\varphi^{[\ell]}_n(\alpha,z)$:
\begin{equation}
	\varphi^{[\ell]}_n(\bar\alpha,\bar z)=
	  \int\limits_{\alpha_{\rm th}}^{\infty}d\alpha
	\int\limits_{-1}^{1}dz \,
	{\cal K}^{[\ell]}_n(\bar\alpha,\bar z;\alpha,z)
	\varphi^{[\ell]}_n(\alpha,z).
	\label{eqn_wght}
\end{equation}
Here we have introduced the total kernel function
${\cal K}^{[\ell]}_n(\bar\alpha,\bar z;\alpha,z)$,
which is the superposition of
$K^{[\ell]}_n(\bar\alpha,\bar z;\alpha,z)$ with the kernel weight functions
$\rho_{\rm ch}(\gamma,\vec\xi)$ such that
\begin{equation}
	{\cal K}^{[\ell]}_n(\bar\alpha,\bar z;\alpha,z)\equiv
	  \sum_{\rm ch}\int\limits_{0}^{\infty } d\gamma\int\limits_{\Omega}
           d\vec\xi
	  \, \rho_{\rm ch}(\gamma,\vec\xi) \,
	  K^{[\ell]}_n(\bar\alpha,\bar z;\alpha,z).
	\label{totl_krnl}
\end{equation}
Since the weight functions $\rho_{\rm ch}(\gamma,\vec\xi)$
for the scattering kernel are real functions by their construction,
the total kernel function ${\cal K}^{[\ell]}_n(\bar\alpha,\bar z;\alpha,z)$
is real, so that
the Eq.~(\ref{eqn_wght}) is a real integral equation with
two variables $\alpha$ and $z$.  Thus we have transformed the
BS equation, which is a singular integral equation of
complex distributions, into a real integral equation which is
frame-independent.
Once one solves for the BS amplitude weight function, the BS amplitude
can be written down in an arbitrary frame.
This is clearly advantageous for applications of the BS amplitude to
relativistic problems.  As we will see the real weight functions can be
in fact be real distributions.
However, as is evident from Eq.~(\ref{rln_wghtfnc})
if $n$ is chosen sufficiently large these
can be always transformed into arbitrarily smooth functions suitable for
numerical solution.

The kernel function $K^{[\ell]}_n(\bar\alpha,\bar z;\alpha,z)$ for
a fixed parameter set
has the following structure;
\begin{eqnarray}
	K^{[\ell]}_n(\bar\alpha,\bar z;\alpha,z) & =  &
	\delta_{n0}\delta(\bar\alpha) h^{[\ell]}_0(\alpha,z)
	+ n\bar\alpha^{n-1}\theta(\bar\alpha)h^{[\ell]}_n(\alpha,z)\nonumber\\
	& & -n\bar\alpha^{n-1} k^{[\ell]}_n(\bar\alpha,\bar z;\alpha,z)
		- g^{[\ell]}_n(\bar\alpha,\bar z;\alpha,z).
	\label{krnl_fixed}
\end{eqnarray}
The functions $h^{[\ell]}_n(\alpha,z)$,
$k^{[\ell]}_n(\bar\alpha,\bar z;\alpha,z)$ and
$g^{[\ell]}_n(\bar\alpha,\bar z;\alpha,z)$ are
defined as Feynman parameter integrals in Appendix \ref{apdx_c}.
The terms containing the function $h^{[\ell]}_n(\alpha,z)$
are independent of $\bar z$.
Since this feature is independent of the kernel parameters,
the total kernel function
${\cal K}^{[\ell]}_n(\bar\alpha,\bar z;\alpha,z)$
also has this structure for any input scattering kernel.
For example, the total kernel function with $n=0$ can be written as
\begin{equation}
	{\cal K}^{[\ell]}_0(\bar\alpha,\bar z;\alpha,z)
	=\delta(\bar\alpha) {\cal H}_0^{[\ell]}(\alpha,z)
		-{\cal G}_0^{[\ell]}(\bar\alpha,\bar z;\alpha,z),
	\label{total_krnl_0}
\end{equation}
where ${\cal H}_0^{[\ell]}(\alpha,z)$ and
${\cal G}_0^{[\ell]}(\bar\alpha,\bar z;\alpha,z)$ are
the kernel-parameter integrals of
the functions $h^{[\ell]}_n(\alpha,z)$ and
$g^{[\ell]}_n(\bar\alpha,\bar z;\alpha,z)$ with the kernel
weight function in Eq.~(\ref{totl_krnl}), respectively.
This structure suggests that the weight function
$\varphi^{[\ell]}_0(\alpha,z)$ contains a $\delta$-function
\begin{equation}
	\varphi^{[\ell]}_0(\alpha,z)
	= \, c^\prime \delta(\alpha) - \tilde \varphi^{[\ell]}_0(\alpha,z),
	\label{phi_0_structure}
\end{equation}
where $c^\prime$ is a constant and $\tilde \varphi^{[\ell]}_n(\alpha,z)$
is a function determined by the kernel.  If one substitutes this form into the
PTIR for the BS amplitude Eq.~(\ref{BS_PTIR}), the $\delta$-function term
gives the product of two propagators with the weight $c^\prime/2$.
As discussed in Section \ref{sec_PTIR_BS}, this is just
a consequence of the fact that the BS amplitude contains
the kinematical singularity due to the two
constituent propagators.  Thus this part of the structure
is totally independent of the details of the scattering kernel,
(i.e, independent of the scattering kernel weight functions).

Now, if the function ${\cal G}_0^{[\ell]}(\bar\alpha,\bar z;\alpha,z)$
in Eq.~(\ref{total_krnl_0}) vanishes in some region
of $\bar\alpha$ around the point $\bar\alpha=0$,
i.e., if the $\delta$-function singularity corresponds to an isolated ``pole''
contribution to the spectrum of the BS amplitude, one can write the
homogeneous integral equation Eq.~(\ref{eqn_wght})
as the following coupled inhomogeneous equations;
\begin{eqnarray}
	 &  & {c^\prime \over \lambda} =
	  c^\prime\int\limits_{-1}^{1}dz \, {\cal H}_0^{[\ell]}(\alpha=0,z)
	  - \int\limits_{\alpha_{\rm th}}^{\infty}\,
		d\alpha\,\int\limits_{-1}^{1}dz \,
	      {\cal H}_0^{[\ell]}(\alpha,z)
	\tilde \varphi^{[\ell]}_0(\alpha,z),
	\label{inhomo_eqn_1}\\
	 &  & {1 \over \lambda}\tilde \varphi^{[\ell]}_0(\alpha,z) =
	   c^\prime\int\limits_{-1}^{1}dz
	     {\cal G}_0^{[\ell]}(\bar\alpha,\bar z;\alpha=0,z)\,+\,
	   \int\limits_{\alpha_{\rm th}}^{\infty}\,d\alpha\,
             \int\limits_{-1}^{1}dz \,
	     {\cal G}_0^{[\ell]}(\bar\alpha,\bar z;\alpha,z)
	        \tilde \varphi^{[\ell]}_0(\alpha,z),
	\label{inhomo_eqn_2}
\end{eqnarray}
where we have introduced an ``eigenvalue'' parameter $\lambda$.  Instead of
solving for the mass of the bound state for a given scattering kernel,
it is more convenient to solve the BS equation as an ``eigenvalue'' problem
for a fixed bound-state mass parameter $P^2$.  We thus solve for the
eigenvalue  $\lambda$ as a function of $P^2$ and the points $P^2$ which give
$\lambda(P^2)=1$  correspond to the masses of the bound state.
For the weight function $\varphi^{[\ell]}_0(\alpha,z)$
containing the $\delta$-function,
(i.e., the weight function Eq.~(\ref{phi_0_structure})
with non-zero $c^\prime$), one can rescale the weight function such that
the strength of the $\delta$-function term is unity.  Hence, one can study
the bound state problem numerically by
iteration {\it even} for the $n=0$ case.  Consider the following formal
expression obtained by iterating the kernel
\begin{equation}
	\varphi^{[\ell]}_0(\bar\alpha,\bar z)=\delta(\bar\alpha)
	      -\int\limits_{-1}^{1}dz\,
		    {\cal G}_0^{[\ell]}(\bar\alpha,\bar z;\alpha=0,z)
                    +\cdots\;.
\end{equation}
There is similarly an expansion for the eigenvalue $\lambda$
\begin{equation}
	 {1 \over \lambda} =\int\limits_{-1}^{1}dz \, {\cal H}_0^{[\ell]}
          (\alpha=0,z)
	  + \int\limits_{\alpha_{\rm th}}^{\infty}\,d\alpha\,
              \int\limits_{-1}^{1}dz \,
	      {\cal H}_0^{[\ell]}(\alpha,z) \,\int\limits_{-1}^{1}dz^\prime \,
	      {\cal G}_0^{[\ell]}(\alpha,z;\alpha^\prime=0,z^\prime)
              +\cdots\;,
\end{equation}
which is the Fredholm series.
The scalar-scalar-ladder model has been formally solved by Sato by means of
the Fredholm solution.  He showed that the second iterated kernel
is bounded and that the Fredholm theorem is applicable for the resulting
regular equation\cite{Sato}.
Nakanishi extended his results
to arbitrary partial wave solutions\cite{Nakanishi63}.

To apply rigorous mathematical theorems, such as Fredholm theory,
it is necessary to know in detail the singularity
structure of the kernel function
${\cal K}^{[\ell]}_n(\bar\alpha,\bar z;\alpha,z)$
for given weight functions $\rho_{\rm ch}(\gamma,\vec\xi)$.
Hence, it is very difficult to discuss their applicability
for the most general form of the scattering kernel.
Instead of considering such theorems, we shall take the more pragmatic path
of establishing whether or not the integral equation Eq.~(\ref{eqn_wght})
is numerically tractable.

\section{Structure of the Kernel Function}\label{Singularity}

In this section we consider the structure of the kernel function
given in Eq.~(\ref{totl_krnl}).
We first consider the possible singularities of the kernel function
$K^{[\ell]}_n(\bar\alpha,\bar z;\alpha,z)$ for arbitrary $n$ with a fixed
kernel parameter set $(\gamma, \vec\xi)$, i.e., for constant
$\{\gamma,a_{\rm ch},b_{\rm ch},c_{\rm ch},\dots,f_{\rm ch}\}$.
We will in this section omit for brevity
the subscript {\rm ch}.
Apart from the trivial $\delta$-function for $n=0$, which becomes a
step-function for $n=1$, the
possible singularities of $K^{[\ell]}_n(\bar\alpha,\bar z;\alpha,z)$ are
those of the functions $h^{[\ell]}_n(\alpha,z)$,
$k^{[\ell]}_n(\bar\alpha,\bar z;\alpha,z)$, and
$g^{[\ell]}_n(\bar\alpha,\bar z;\alpha,z)$.

Let us first consider the function $g^{[\ell]}_n(\bar\alpha,\bar z;\alpha,z)$.
The integration over the Feynman parameter is easily performed
due to the $\delta$-function.
Reflecting the integral range of the Feynman
parameter $y \in [0, \infty)$, we see that
$g^{[\ell]}_n(\bar\alpha,\bar z;\alpha,z)$ has finite
support as a function of $\bar\alpha$, $\alpha$, $\bar z$ and $z$.
The function $g^{[\ell]}_n(\bar\alpha,\bar z;\alpha,z)$ becomes after the
Feynman parameter integral
\begin{eqnarray}
	& &g^{[\ell]}_n(\bar\alpha,\bar z;\alpha,z) \label{g_krnl}\nonumber\\
	& &= \frac{1}{(4\pi)^2}
	\text{Pf}\,\cdot {1\over \bar\alpha}\left(-\frac{b}{2}\right)^\ell
		\frac{(1-\bar z^2)^{n+1}}{2}
	\sum\limits_{\kappa=\pm 1}\,\text{Pf}\,\cdot \frac{
			 \theta(\tilde B_\kappa(\alpha,z; \bar\alpha,\bar z)^2
			 -4 \tilde A(\alpha,z; \bar z)
			 \tilde C_\kappa(\bar\alpha,\bar z))
		}{\left(
			 \tilde B_\kappa(\alpha,z; \bar\alpha,\bar z)^2
			 -4 \tilde A(\alpha,z; \bar z)
			 \tilde C_\kappa(\bar\alpha,\bar z)
		\right)^{1/2}}
	\label{exp_g_func}\\
	& & \;\;\times\sum\limits_{i=1,2}\,
		\frac{
			{y^\kappa_i}^{n+1}\, (a+y^\kappa_i)^{n-1-\ell}
		}{
			\left(\zeta_\kappa(\bar z,z) y^\kappa_i
				 + \eta_\kappa(\bar z) \right)^n
		}
		\theta(y^\kappa_i)
		\theta\left(\kappa
			\left[ \left(c\bar z -f + {b\over 2}\right)
                        (y^\kappa_i+a)
			- {b \over 2} \left(\bar z\, {b\over 2}-(e-a z) \right)
			\right]
		\right)\;,
	\nonumber
\end{eqnarray}
where the symbol $\text{Pf}\,\cdot$ stands for the Hadamard finite part.
Any singularities that arise from the Feynman parameter
integrals should be regularized using the Hadamard finite part prescription.
This is consistent with the ordinary $i\epsilon$ prescription
for a calculation of Feynman diagrams in momentum space
\cite{Nakanishi_graph,Nakanishi63}.
A discussion of the finite parts of
singular integrals and the detailed calculation
of $g^{[\ell]}_n$ are given in Appendix \ref{apdx_d}.
The variables $y^\kappa_i$ for $i=1,2$ are the roots of
the equation $\tilde A(\alpha,z; \bar z) y^2
+ \tilde B_\kappa(\alpha,z; \bar\alpha,\bar z) y
+ \tilde C_\kappa(\bar\alpha,\bar z) = 0$.
See Appendix \ref{apdx_c} for the
definition of the functions $\tilde A(\alpha,z; \bar z)$,
$\tilde B_\kappa(\alpha,z; \bar\alpha,\bar z)$,
$\tilde C_\kappa(\bar\alpha,\bar z)$, $\zeta_\kappa(\bar z,z)$
and $\eta_\kappa(\bar z)$.
Since the factor $\zeta_\kappa(\bar z,z) y^\kappa_i
 + \eta_\kappa(\bar z)$ in the denominator of Eq.~(\ref{exp_g_func})
is positive definite for any positive $y^\kappa_i$,
this factor does not cause any singularity.  Thus
for $\tilde C_\kappa(\bar\alpha,\bar z) < 0$,
$g^{[\ell]}_n(\bar\alpha,\bar z;\alpha,z)$ is regular everywhere
except the pole at $\bar\alpha = 0$, since $\tilde A(\alpha,z; \bar z)$ is
non-negative for any $\alpha$, $z$ and $\bar z$.
For $\tilde C_\kappa(\bar\alpha,\bar z) \ge 0$ the square root of the
following factor in the denominator can vanish
\begin{eqnarray}
	& &\tilde B_\kappa(\alpha,z; \bar\alpha,\bar z)^2
			     -4 \tilde A(\alpha,z; \bar z)
			     \tilde C_\kappa(\bar\alpha,\bar z)
	\nonumber\\
	& & \quad = \frac{1}{a^2}
	    \left(a^2 \tilde A(\alpha,z;\bar z)
			  -\left(
			    \sqrt{\tilde D_\kappa(\bar\alpha,\bar z, z)}
			    -\sqrt{\tilde C_\kappa(\bar\alpha,\bar z) }
			  \right)^2
		  \right)
     	\label{snglrty}\\
	& & \qquad\quad\times
		\left(a^2 \tilde A(\alpha,z; \bar z)
		    -\left( \sqrt{\tilde D_\kappa(\bar\alpha,\bar z, z)}
			           +\sqrt{\tilde C_\kappa(\bar\alpha,\bar z)}
			\right)^2
		\right),
	\nonumber
\end{eqnarray}
where $\tilde D_\kappa(\bar\alpha,\bar z, z)$ is the regular function
defined in Appendix \ref{apdx_c}.
This singularity occurs at the boundary of the step function.
Since a square root singularity is integrable, this singularity
is numerically tractable for
$\tilde C_\kappa(\bar\alpha,\bar z) > 0$.
In the general case one can apply the Hadamard finite part prescription
(\ref{Hadamard's_orgn}) in Appendix \ref{apdx_d}
for a regularization.  However, when two roots coincide, namely
at the point $(\bar\alpha, \bar z)$
on which $\tilde C_\kappa(\bar\alpha,\bar z) = 0$,
special care is necessary to perform the $\alpha$-integral.
One may handle this singularity by the regularized expression
Eq.~(\ref{exp_reg_g_func}) with the finite regularization
parameter $\epsilon$.
Thus $\varphi^{[\ell]}_n(\alpha,z)$ may have singularities even
for a constant kernel parameter set in the general case.
Note that this singular structure of
$g^{[\ell]}_n(\bar\alpha,\bar z;\alpha,z)$
is independent of $n$ and $l$.

Now let us turn to the functions $h^{[\ell]}_n(\alpha,z)$ and
$k^{[\ell]}_n(\bar\alpha,\bar z;\alpha,z)$.  Since the step function
in $k^{[\ell]}_n(\bar\alpha,\bar z;\alpha,z)$ restricts the integral
range of $y$ for a given parameter set, it is enough to consider
the integral
\begin{equation}
	I^{[\ell]}_n(y_{\rm min},y_{\rm max})\equiv
	\int_{y_{\rm min}}^{y_{\rm max}}dy\,\frac{y^{n+1}(a+y)^{n-1-l}}
		{[A(\alpha,z) y^2 + B(\alpha,z) y +C]^{n+1}}
	\label{func}
\end{equation}
in order to discuss the singularity structure.
It is easy to show that Eq.~(\ref{func}) can be written as
\begin{equation}
	I^{[\ell]}_n(y_{\rm min},y_{\rm max})
		\left.
		={(-1)^{n+\ell} \over n! \ell!}
		\left(\frac{\partial}{\partial \alpha}\right)^n
		\left(\frac{\partial}{\partial \tilde a}\right)^\ell
		\tilde I(y_{\rm min},y_{\rm max}) \right|_{\tilde a = a}\;,
	\label{func2}
\end{equation}
with the function
\begin{equation}
	\tilde I(y_{\rm min},y_{\rm max})\equiv
	\int_{y_{\rm min}}^{y_{\rm max}}dy\,\frac{y}{\tilde a+y}
		\frac{1}{[A(\alpha,z) y^2 + B(\alpha,z) y +C]^{n+1}}\;.
	\label{func0}
\end{equation}
The boundary of the integral $y_{\rm min}$ and $y_{\rm max}$ should be
understood as constant, when taking
the derivative with respect to $\alpha$ and $\tilde a$, although they are
fixed as functions of variables by the step function of Eq.~(\ref{step_y}).
Since the $y$-integral should be performed using the Hadamard finite part
prescription, the above operation is valid even though
$I^{[\ell]}_n(y_{\rm min},y_{\rm max})$ contains singularities.
The integral range $[y_{\rm min}, y_{\rm max}]$ for
$k^{[\ell]}_n(\bar\alpha,\bar z;\alpha,z)$
depends on
$\bar\alpha$, $\bar z$, $\alpha$ and $z$ and for
$h^{[\ell]}_n(\alpha,z)$ is fixed to $[0, \infty)$.
Since $A(\alpha,z) > 0$ for
any $z$ and $\alpha$ that is consistent with the support
property of the weight function Eq.~(\ref{Supp_BS_PTIR}),
the integrand always vanishes sufficiently fast for $y \rightarrow\infty$.
Thus a singularity occurs only if the set of variables; $\bar\alpha$,
$\bar z$, $\alpha$ and $z$, satisfy the condition where the
denominator of the integrand $A y^2 + B y +C$ vanishes for some $y$,
namely $B^2-4AC \ge 0$ \footnote{
$I^{[\ell]}_n(y_{\rm min}=0,y_{\rm max})$ at $a=0$ can be singular.  Such
a singular
point is independent of the choice of the variables.  One can always
avoid it by choosing the dummy parameter to be some $n \ge l+1$.}.

For $B^2-4AC \ge 0$, the integration over $y$ yields
\begin{eqnarray}
	& \tilde I(y_{\rm min},y_{\rm max}) =
	\frac{1}{\tilde a^2 A -\tilde a B + C}
	&\left\{ \lim_{\epsilon\rightarrow 0}
		\frac{\tilde a}{8}\,
		\ln\left({Y^2+\epsilon^2}\right)
	\right. \nonumber\\
	& &\qquad \left. \left.
	+\lim_{\epsilon\rightarrow 0}
	\frac{2C-\tilde a B}{ 8\sqrt{B^2-4AC} }
		\ln\left(\frac{(1+X)^2+\epsilon^2}
						{(1-X)^2+\epsilon^2}\right)
	\right\} \right|_{y_{\rm min}}^{y_{\rm max}}\;,
	\label{int_func0}
\end{eqnarray}
with $X=\frac{\sqrt{B^2-4AC}\, y}{2C+By}$ and
$Y=\left(\frac{\tilde a}{y+\tilde a}\right)^2(Ay^2+By+C)$.
As shown in Appendix \ref{apdx_c} the factor
$\tilde a^2 A -\tilde a B + C$ becomes a positive definite
function $D(z)$ depending only on $z$ after setting $\tilde a = a$.
The limit $B^2-4AC \rightarrow 0$ is also regular, since $X$ is
proportional to the factor $\sqrt{B^2-4AC}$.
Thus a singularity occurs
only when $1\pm X$ or $Y$ vanish at the end point of the $y$-integration.
Furthermore, neither the derivative with respect to $\alpha$
nor that with respect to $\tilde a$ generates any new singular point.
They simply change the power of the singular behavior.  Since $X$ is
independent of $\tilde a$ and ${\partial Y\over \partial\tilde a}
=2{y\over y+\tilde a}Y$, the derivative with respect to $\tilde a$
acting on $\tilde I(y_{\rm min},y_{\rm max})$ does not cause stronger
singularities.
On the other hand, the derivative with respect to $\alpha$
increases the inverse power of $Y$ and $1\pm X$.  Thus the possible
singularity of the integral $I^{[\ell]}_n(y_{\rm min},y_{\rm max})$ is of
the form
\begin{equation}
	\text{Pf}\,\cdot
	   \frac{1}{\left( A y_{\rm min}^2 + B y_{\rm min} + C \right)^m }\quad
	   \text{or}\quad
	 \text{Pf}\,\cdot
	  \frac{1}{\left( A y_{\rm max}^2 + B y_{\rm max} + C \right)^m },
	\label{snglrty2}
\end{equation}
where $m$ is a non-negative integer which does not exceed $n$ and where we have
used the fact that $1/(1\pm X)\propto 1/(1-X^2)\propto 1/(A y^2 + B y + C)$.
It is not difficult to see that the fixed end points ($y=0$ and $y=\infty$),
which are independent of the choice of parameters, cause no
further singularities provided $B(\alpha,z)$ or $C$ does not vanish.
Thus the function $h^{[\ell]}_n(\alpha,z)$ has no singularity,
if $C\ne 0$.
On the other hand, $k^{[\ell]}_n(\bar\alpha,\bar z;\alpha,z)$
may have singularities.
A possible end point of the $y$-integral,
which might give the singularity, is;
the positive root $y^\kappa_i$, $(i=1,2)$ of the equation
\begin{equation}
	\tilde A(\alpha,z; \bar z)y^2
		+\tilde B_\kappa(\alpha,z; \bar\alpha,\bar z) y
		+\tilde C_\kappa(\bar\alpha,\bar z) = 0,
	\label{bndry_eqn0}
\end{equation}
and the boundary value of $y$ at which $\kappa$ changes sign:
$y_{\text{bd}} =-a+ {b\over 2} \frac{\bar z\, b/2-e+a z}{c\bar z -f + b/2}$
when $y_{\text{bd}} \ge 0$.  Since the Eq.~(\ref{bndry_eqn0}) can be written
as follows
\begin{equation}
	A(\alpha,z)y^2+B(\alpha,z) y + C =
		\frac{\zeta_\kappa(\bar z,z)y + \eta_\kappa(\bar z)}
			{1-\bar z^2}\bar\alpha,
\end{equation}
and the factor $\zeta_\kappa(\bar z,z)y + \eta_\kappa(\bar z)$ is positive
definite, the end point singularity at $y=y^\kappa_i$ is given by
$\text{Pf}\,\cdot 1/\bar\alpha^n$.
Thus the possible singuralities of $k^{[\ell]}_n(\bar\alpha,\bar z;\alpha,z)$
are $\text{Pf}\,\cdot 1/\bar\alpha^n$ and
$\text{Pf}\,\cdot 1/(A y_{\text{bd}}^2+B y_{\text{bd}}+C)^m$ with $0 \le m \le
n$,
where $n, m=0$ stands for logarithm.
However, $k^{[\ell]}_n(\bar\alpha,\bar z;\alpha,z)$ appears in
$K^{[\ell]}_n(\bar\alpha,\bar z;\alpha,z)$
together with the additional factor $n\bar\alpha^{n-1}$,
so that $k^{[\ell]}_n(\bar\alpha,\bar z;\alpha,z)$ exists only for
$n \ge 1$.  The logarithmic singularities then disappear
in $K^{[\ell]}_n(\bar\alpha,\bar z;\alpha,z)$, because
the derivative with respect to $\alpha$ cancel the logarithmic term
in (\ref{int_func0}).  Furthermore the factor $\bar\alpha^{n-1}$
weakens the singularity at $\bar\alpha=0$, so that only
the singularity $\text{Pf}\,\cdot 1/\bar\alpha$
survives for $n\ge 1$ and any $l$.

To summarize, the possible singularities of the
kernel function $K^{[\ell]}_n(\bar\alpha,\bar z;\alpha,z)$ for
a fixed kernel parameter set $\{a_{\rm ch},b_{\rm ch},c_{\rm ch},\dots,
f_{\rm ch}\}$
are: a trivial
$\delta$-function singularity at $\bar\alpha=0$ for the choice of
dummy parameter $n=0$;
$\text{Pf}\,\cdot 1/\bar\alpha$ for $n\ge 1$;
$\text{Pf}\,\cdot 1/(A y_{\text{bd}}^2+B y_{\text{bd}}+C)^m$
with $1 \le m \le n$ for $n\ge 1$;
the singularity of Eq.~(\ref{snglrty}) at the boundary of
the support for all $n$ and $l$.

\section{Numerical Results}
\label{Results}

In this section we present numerical solutions for the BS
amplitude for bound states in scalar theories using
Eq.(\ref{eqn_wght}) for
two simple choices of scattering kernels:
(a) pure ladder kernel with massive scalar exchange, and
(b) a generalized kernel combined with the above pure ladder
kernel.

The scattering kernel (a), i.e., the one-$\sigma$-exchange
kernel depending only on $t=(p-q)^2$, is given by
Eq.({\ref{pure_ladder_kernel}}).
This corresponds to choosing for the kernel in Eq.~(\ref{krnl_PTIR}) say:
$\rho_{tu}=\rho_{us}=0$ and in the $st$-channel $\gamma=m_\sigma^2$,
and $a_{st}=c_{st}=1$, $b_{st}=-2$, $d_{st}=e_{st}=f_{st}=0$,
[c.f., Eq.~(\ref{krnl_wght})].
This corresponds to choosing $\rho_{st}$ to be some appropriate product of
$\delta$-functions multiplied by $g^2$.
In the pure ladder case it is
convenient (and traditional) to factorize out
the coupling constant $g^2$ and a factor of $(4\pi)^2$, by defining
the ``eigenvalue'' $\lambda=g^2/(4\pi)^2$ \cite{Sato,Nakanishi63,L+M}.
Thus it is usual to fix the bound state mass $P^2$ and then to solve for
the coupling $g^2$, which is what we have done here.
In the numerical calculations it is necessary to regularize the integrable
square root singularity discussed in the previous section
(see Eq.(\ref{snglrty})), i.e., we kept the regularization parameter
$\epsilon$ in the Hadamard finite part Pf$\cdot 1/x^n$
small but finite, typically $\sim 10^{-5}$ or less.  Of course, we verified
that our solutions were independent of $\epsilon$ provided it was chosen
sufficiently small.  We began by discretizing the $\alpha$ and $z$ axes
and then solved the integral equation by iteration from some initial
assumed weight function.  It was also, of course, confirmed that the solution
was robustly independent of the choice of
starting weight function, the number
of grid points, and the maximum grid value of $\alpha$ (when the latter
two of these were chosen suitably large).
We have solved over a range of $P^2$ and $m_\sigma$ and all solutions
reproduced the eigenvalues obtained in Euclidean space
after a Wick rotation by Linden and Mitter \cite{L+M}.
We found that this simple method produces convergence in
typically 10 cycles or so and for the relatively modest grid choices used
gave results which agreed to within approximately 1\% of the Wick-rotated
ones.  In Fig. (\ref{fig_pure}) we plot the weight
function $\phi_{n=1}(\alpha,z)$ for the bound state with mass given by
$P^2/m^2=(M/m)^2=3.24$, where $m$ is the mass of
the constituent particles and
where the mass of the exchanged particle is $m_\sigma/m=0.5$.
This solution was generated by solving for $n=0$ and then using
Eq.~(\ref{rln_wghtfnc}) to convert to the $n=1$ form of the solution.
This has the advantage of filtering out numerical noise and the
resulting solution is more readily graphically represented.
The coupling
constant for this case was found to be $g^2/(4\pi)^2=1.02$, c.f., the Linden
and Mitter result of 1.0349.

For a nontrivial example of the scattering kernel, where the na\"{\i}ve
Wick-rotation is not possible,
we studied the following simple generalization of the pure ladder case:
a sum of the one-$\sigma$-exchange kernel (a)
and the generalized kernel term with the following fixed
kernel parameter set
\begin{eqnarray}
	& &a_{\rm st}= 0.47261150181,\nonumber\\
	& &b_{\rm st}=-0.29743163287,\nonumber\\
	& &c_{\rm st}= 0.58277042955,\nonumber\\
	& &d_{\rm st}= 0.28282145969,\nonumber\\
	& &e_{\rm st}=-0.23965580016,\nonumber\\
	& &f_{\rm st}= 0.32196629047,\nonumber\\
	& &\gamma/m^2= 2.25,
	\label{paras}
\end{eqnarray}
weighted with a factor $(1/4)[g^2/(4\pi)^2]$.  In addition, the resulting
total kernel was
symmetrized in $\bar z \rightarrow -\bar z$ so that we obtain
appropriate symmetry for $s$-wave normal solutions.
We obtained these parameters by generating $\xi_i$ in the $st$-channel
with a random number generator.
We then solved for $\lambda=g^2/(4\pi)^2$ as for the pure ladder
case but for this case the $n=1$ was solved for directly, without first
obtaining the $n=0$ solution.
In Fig. (\ref{fig_gnrl}) we plot the weight function
$\phi_{n=1}(\alpha,z)$ with the mass parameters again chosen
as $P^2/m^2=3.24$ and $m_\sigma/m=0.5$.
Some additional structure is apparent, as expected,
and some residual numerical noise is also present in this figure.
Methods of reducing this numerical noise by finding more
sophisticated choices of grid and integration method are being
investigated~\cite{KandW}. The resulting eigenvalue
is $\lambda=0.889$, which shows that the addition to the pure
ladder kernel has enhanced the binding and so is attractive.
We found that the convergence and the stability of the eigenvalues
with varying the number of grid points, $\epsilon$, and $\alpha_{\rm max}$
are excellent as for the pure ladder case above.

\section{Summary and Conclusions}
\label{Conclusions}

We have derived a real integral equation for the weight function of the
scalar-scalar Bethe-Salpeter (BS) amplitude from the BS equation for
scalar theories without derivative coupling.
This was achieved using the perturbation theory integral representation
(PTIR), which is an extension of the spectral representation for two-point
Green's functions, for both the scattering kernel [Eq.~(\ref{krnl_PTIR})]
and the BS amplitude itself [Eqs.~(\ref{ptlw_BS_PTIR},\ref{ptlw_BS_PTIR2})].
The uniqueness theorem of the PTIR and the appropriate application
of Feynman parameterization then led to the central result of the paper
given in Eq.~(\ref{eqn_wght}).
We demonstrated that this integral equation is numerically tractable
for both the pure ladder case and an arbitrary generalization of this.
We have verified that our numerical solutions agree with those previously
obtained from a Euclidean treatment of the pure ladder limit.

This represents a potentially powerful new approach to obtaining solutions of
the BS equation and additional results and applications are currently being
investigated \cite{KandW}.  The separable kernel case should be studied,
since it has exact solutions and so is a further independent
test of the approach
developed here.  While our detailed discussions were limited to the equal mass
case, it is worthwhile to investigate generalizations including
nonperturbative constituent propagators and the heavy-light bound state
limit to see under which conditions an approximate Klein-Gordon equation
can result.
It is also important to find a means to generalize
this approach to include derivative couplings and fermions.

\begin{acknowledgments}

This work was supported by the Australian Research Council and
also in part by grants of
supercomputer time from the U.S. National Energy Research Supercomputer
Center and the Australian National University
Supercomputer Facility.

\end{acknowledgments}

\appendix

\section{PTIR for Scattering Kernel}\label{apdx_a}

In this appendix we list the dimensionless coefficients
$\{a_{\rm ch},b_{\rm ch},c_{\rm ch},\dots,f_{\rm ch}\}$
in Eq. (\ref{krnl_wght})
for different channels {ch}$=\{{st}\},\{{tu}\},\{{us}\}$
in terms of the Feynman parameters $\xi_i$ defined in Eq. (\ref{krnl_PTIR}).

\begin{center}
\begin{tabular}{cccc}
	\hline
	 & $st$ & $tu$ & $us$  \\
	\hline\hline
	$\quad a_{\rm ch}\quad$
& $\xi_1+\xi_2+\xi_6$ & $\xi_1+\xi_2+\xi_5+\xi_6$ & $\xi_1+\xi_2+\xi_5$  \\
	$b_{\rm ch}$ & $-2 \xi_6$ & $2(\xi_6-\xi_5)$ & $2\xi_5$  \\
	$c_{\rm ch}$ & $\xi_3+\xi_4+\xi_6$ & $\xi_3+\xi_4+\xi_5+\xi_6$ &
$\xi_3+\xi_4+\xi_5$  \\
	$d_{\rm ch}$ & $\quad{1\over 4}(\xi_1+\xi_2+\xi_3+\xi_4)+\xi_5\quad$
& $\quad{1\over 4}(\xi_1+\xi_2+\xi_3+\xi_4)\quad$ &
	    $\quad{1\over 4}(\xi_1+\xi_2+\xi_3+\xi_4)+\xi_6\quad$  \\
	$e_{\rm ch}$ & $\xi_1-\xi_2$ & $\xi_1-\xi_2$ & $\xi_1-\xi_2$  \\
	$f_{\rm ch}$ & $\xi_3-\xi_4$ & $\xi_3-\xi_4$ & $\xi_3-\xi_4$  \\
	\hline
\end{tabular}\label{table_abc}
\end{center}

\section{Solid Harmonics}\label{apdx_b}

Let us consider the integral
\begin{eqnarray}
	I(n,m,\ell;p,P)&\equiv&
		\int\frac{d^4q}{(2\pi)^4i}
		\frac{1}
			{\left[\gamma-
			\left(
				a\,q^2+b\,p\cdot q+c\,p^2+d\,P^2
				+e\,q\cdot P+f\,p\cdot P
			\right)-i\epsilon\right]^m}\nonumber\\
	& & \quad\times\frac{{\cal Y}_\ell^{\ell_z}(\Lambda^{-1}(P) q)}
			{\left[m^2+\alpha-\left(q^2 + z q\cdot P + P^2/4\right)
			-i\epsilon\right]^{n}}
	\label{mom_int}
\end{eqnarray}
Applying the Feynman parameterization we have
\begin{eqnarray}
	& & I(n,m,\ell;p,P)\nonumber\\
	& = &
		\int\frac{d^4q}{(2\pi)^4i}
		\frac{\Gamma(n+m)}{\Gamma(n)\Gamma(m)}
			\int\limits_0^1 dx
		\frac{x^{n-1}(1-x)^{m-1}}{(x+(1-x)a)^{n+m}}\\
	& &\times\frac{{\cal Y}_l^{l_z}(\Lambda^{-1}(P) q)}
			{\left[{x(1-x)\over (x+(1-x)a)^2}H\left[
			{F\over H} + m^2
			-\left(p^2+{P^2\over 4}+J p\cdot P\right)\right]
				-\left(q+\frac{(xz+(1-x)e)P+(1-x)b p}
				{2(x+(1-x)a)}\right)^2-i\epsilon
			\right]^{n+m}},\nonumber
\end{eqnarray}
where $F$, $H$ and $J$ are
\begin{eqnarray}
	& & F=A(\alpha,z)\frac{x}{1-x}+B(\alpha,z)+C\frac{1-x}{x},\nonumber\\
	& & H=c+\Delta \frac{1-x}{x},\nonumber\\
	& & J=f-b/2 z + \frac{1-x}{x}(af-e{b\over 2}),\nonumber
\end{eqnarray}
and $A(\alpha,z)$, $B(\alpha,z)$ and $C$ are defined
in Appendix \ref{apdx_c}.
Introducing a new variable
$q'=\Lambda^{-1}(P)\left(q+\frac{(xz+(1-x)e)P+(1-x)bp}{2(x+(1-x)a)}\right)$
and recalling the fact $\Lambda^{-1}(P)P=(\sqrt{P^2},\vec 0)$,
one can factorize the solid harmonics
outside the integral by making use of the
following property of the solid harmonics;
\begin{equation}
	\int d^3q F(\vec q\,^2)
		{\cal Y}_\ell^{\ell_z}(\vec q+\vec p)
	={\cal Y}_\ell^{\ell_z}(\vec p)\int d^3q
		F(\vec q\,^2),
  \label{solid_harm_prop}
\end{equation}
where $F$ is a sufficiently rapidly decreasing function which
gives a finite integral \cite{Nakanishi63}.
The integration over $q$ then yields
\begin{eqnarray}
	& & I(n,m,\ell;p,P) \nonumber \\
	& = &	{\cal Y}_\ell^{\ell_z}(\Lambda^{-1}(P) p)
		\frac{ \Gamma(n+m-2) }{ \Gamma(n) \Gamma(m) }
		\frac{1}{ (4\pi)^2 }
		\left( -{b \over 2} \right)^\ell
			\int\limits_0^\infty dy
		\frac{y^{n-1} (a+y)^{n+m-4-\ell} }
			{ [c y +\Delta]^{n+m-2} }  \nonumber \\
	&  & \quad \times \frac{1}
			{\left[\frac{A(\alpha,z)y^2+B(\alpha,z)y+C}
				{c y +\Delta}
				+ m^2
				-\left(p^2+{P^2\over 4}+J p\cdot P\right)
				-i\epsilon
			\right]^{n+m-2}}.
\end{eqnarray}
where we have introduced a new integration variable $y$
such that $y=x/(1-x)$.  Thus the integration over the loop momentum
reproduces the same solid harmonic.
This self-reproducing property of the solid harmonics ensures
that the integral representation, Eq.~(\ref{ptlw_BS_PTIR2}),
forms an irreducible representation of the Poincare group.

\section{Kernel Function}\label{apdx_c}

In this Appendix we present the explicit expression of the kernel function
for the constant kernel parameter set ${a,b,c,d,e,f}$.
The kernel function $K_n(\bar\alpha,\bar z;\alpha,z)$ for the weight function
$\varphi_n^{[\ell]}(\alpha,z)$ is defined
as the following Feynman parameter integral
\begin{eqnarray}
	K_n^{[\ell]}(\bar\alpha,\bar z;\alpha,z)&=&\frac{1}{(4\pi)^2}
		\frac{1}{2} \left(-{b\over 2}\right)^l\int_{0}^{\infty}dy
		\frac{y^{n+1} (a+y)^{n-1-l}}
		{\left[A(\alpha,z)y^2+B(\alpha,z)y+C\right]^{n+1}}
	\label{gnrl_n_kernel}\\
	& & \, \times \frac{\partial}{\partial \bar\alpha}\bar\alpha^n
	  \left[\theta(\bar\alpha)-\theta\left(\bar\alpha-R(\bar z, G(z))
	  \frac{A(\alpha,z)y^2+B(\alpha,z)y+C}{c y+\Delta}\right)\right],
	\nonumber
\end{eqnarray}
where have suppressed the explicit dependence on the kernel
parameters in $A(\alpha,z)$,$B(\alpha,z)$ and $C$.  These are defined as
\begin{eqnarray}
	& &A(\alpha,z)  =  \alpha + m^2 -(1-z^2){P^2\over 4},
	\nonumber \\
	& &B(\alpha,z)  = a\alpha+\gamma + (a-c)\left(m^2-{P^2\over 4}\right)
				-(4d-2ez){P^2\over 4},
	\label{para_def} \\
	& &C  =  a \gamma - \Delta\left(m^2-{P^2\over 4}\right)
			-(4ad-e^2){P^2\over 4},
	\nonumber
\end{eqnarray}
where $\Delta\equiv ac-b^2/4$.
The functions $R(\bar z,z)$ and $G(z;y)$ are defined as
\begin{eqnarray}
	& &R(\bar z,z) \equiv  \frac{1-\bar z}{1- z}\theta(\bar z - z) +
	                  \frac{1+\bar z}{1+ z}\theta(z - \bar z ),
	\label{R_func}\\
	& &G(z;y)  \equiv  \frac{(f-b/2 z)y+af-eb/2}{c y+\Delta}.
	\label{G_func}
\end{eqnarray}
Note that these functions are bounded, such that
\begin{eqnarray}
	& & 0 \le R(\bar z,z) \le 1 \quad \text{for}\quad |\bar z|,|z| \le 1,
	\label{R_bound}\\
	& &|G(z;y)| \le 1 \quad \text{for}\quad
				|z| \le 1 \,\text{and}\, y \ge 0,
	\label{G_bound}
\end{eqnarray}
provided $|f\pm b/2|\le c$ and $|af-eb/2|\le\Delta$.  These conditions are
automatically satisfied, which is readily seen
if one rewrites ${a,b,\dots,f}$ by $\vec\xi \in \Omega$, with
$\Omega\equiv\{\xi_i \,| \,0\leq \xi_i \leq 1, \,\sum\xi_i=1 \,
(i=1,\dots, 6)\}$.
To integrate over $y$, the denominators of the integrand [i.e.,
$y^{n-1-l}$ and $(A(\alpha,z)y^2+B(\alpha,z)y+C)^{-(n+1)}$]
should be understood
as the Hadamard finite part, if necessary.
Since $G(\bar z=0;y)=0$, the kernel function
$K_n^{[\ell]}(\bar\alpha,\bar z;\alpha,z)$ vanish identically
at $\bar z=0$, so that we need only consider the case $\bar z\ne 0$ from
now on.

Performing the derivative with respect to $\bar\alpha$, the kernel function
Eq.~(\ref{gnrl_n_kernel}) becomes
\begin{eqnarray}
	K^{[\ell]}_n(\bar\alpha,\bar z;\alpha,z) & =  &
	\frac{\partial}{\partial\bar\alpha}\left(
	\bar\alpha^{n}\theta(\bar\alpha)\right)h^{[\ell]}_n(\alpha,z)
	\nonumber\\
	& & -\left[
		n\bar\alpha^{n-1} k^{[\ell]}_n(\bar\alpha,\bar z;\alpha,z)
		+ g^{[\ell]}_n(\bar\alpha,\bar z;\alpha,z)
	\right]
	\label{krnl_func}
\end{eqnarray}
with the functions $h^{[\ell]}_n(\alpha,z)$,
$k^{[\ell]}_n(\bar\alpha,\bar z;\alpha,z)$
and $g^{[\ell]}_n(\bar\alpha,\bar z;\alpha,z)$;
\begin{eqnarray}
	& h^{[\ell]}_n(\alpha,z) &=\frac{1}{(4\pi)^2}
		\frac{1}{2}\left(-\frac{b}{2}\right)^l
		\int_{0}^{\infty}dy
		\frac{y^{n+1} (a+y)^{n-1-l}}
	  		{\left[A(\alpha,z)y^2+B(\alpha,z)y+C\right]^{n+1}},
	\label{h_func}\\
	& k^{[\ell]}_n(\bar\alpha,\bar z;\alpha,z) & = \frac{1}{(4\pi)^2}
		\frac{1}{2}\left(-\frac{b}{2}\right)^l
		\int_{0}^{\infty}dy
		\frac{y^{n+1} (a+y)^{n-1-l}}
			{\left[A(\alpha,z)y^2+B(\alpha,z)y+C\right]^{n+1}}
	\nonumber\\
		& & \quad\times
		\theta\left(
			\bar\alpha-R(\bar z, G(z;y))
			\frac{A(\alpha,z)y^2+B(\alpha,z)+C}{c y+\Delta}
		\right),
	\label{k_func}\\
	& g^{[\ell]}_n(\bar\alpha,\bar z;\alpha,z) & = \frac{1}{(4\pi)^2}
		\frac{\bar\alpha^n}{2}\left(-\frac{b}{2}\right)^l
		\int_{0}^{\infty}dy
		\frac{y^{n+1} (a+y)^{n-1-l}}
			{\left[A(\alpha,z)y^2+B(\alpha,z)y+C\right]^{n+1}}
	\nonumber\\
		& & \quad\times
		\delta\left(
			\bar\alpha-R(\bar z, G(z;y))
			\frac{A(\alpha,z)y^2+B(\alpha,z)+C}{c y+\Delta}
		\right).
	\label{g_func}
\end{eqnarray}

Now let us consider the $y$-integration.  We first rewrite the
step function and the delta function as functions of $y$.
Substituting (\ref{R_func}) and (\ref{G_func}), the argument of the
step function and the $\delta$-function can be written as
\begin{eqnarray}
	& &\bar\alpha-R(\bar z, G(z;y))
		\frac{A(\alpha,z)y^2+B(\alpha,z)y+C}
		{c y+\Delta}
		\nonumber\\
	& = &
		-\frac{
			\tilde A(\alpha,z; \bar z)y^2
			+\tilde B_\kappa(\alpha,z; \bar\alpha,\bar z) y
			+\tilde C_\kappa(\bar\alpha,\bar z)
		}
		{\zeta_\kappa(\bar z,z) y+ \eta_\kappa(\bar z)}
		\label{arg_step}
\end{eqnarray}
with $\kappa = \pm1$ for
\begin{equation}
\left(c\bar z -f +{b\over 2}z\right)(y+a)
	- {b\over 2}
	\left(
		{b\over 2}\bar z - (e - a z)
	\right)
	\matrix{>\cr
	<\cr
    } 0
	\label{cnd_kap}
\end{equation}
and $\kappa = 0$ when LHS of (\ref{cnd_kap}) vanishes.
We have introduced the coefficients for the denominator defined as
\begin{eqnarray}
	& &\zeta_\kappa(\bar z,z)=
		\left(c+\kappa\left({b\over 2}z-f\right)\right)
			(1+\kappa \bar z)
		+ (1-\kappa^2)\left({b\over 2}z-f\right)\bar z,\nonumber\\
	& &\eta_\kappa(\bar z)=
		\left(\Delta-\kappa\left(af-{b\over 2}e\right)\right)
			(1+\kappa \bar z)
		+ (1-\kappa^2)\left({b\over 2}e-af\right)\bar z,
\end{eqnarray}
and for the numerator we have
\begin{eqnarray}
	& &\tilde A(\alpha,z; \bar z)=(1-\bar z^2) A(\alpha,z)\;,\nonumber \\
	& &\tilde B_\kappa(\alpha,z; \bar\alpha,\bar z)
		 =  (1-\bar z^2) B(\alpha,z)
			- \zeta_\kappa(\bar z,z) \bar\alpha\;,\nonumber\\
	& &\tilde C_\kappa(\bar\alpha,\bar z)
		  =  (1-\bar z^2)C - \eta_\kappa(\bar z) \bar\alpha\;.
\end{eqnarray}
Note that $\zeta_\kappa(\bar z,z)\ge 0$ and $\eta_\kappa(\bar z)\ge 0$ for
$|\bar z|,|z| \le 1$.  Thus the denominator is non-negative for positive $y$,
so that the sign of the argument (\ref{arg_step}) becomes positive only if
the following equation has a positive root
\begin{equation}
	\tilde A(\alpha,z; \bar z)y^2
		+\tilde B_\kappa(\alpha,z; \bar\alpha,\bar z) y
		+\tilde C_\kappa(\bar\alpha,\bar z) = 0\;,
	\label{bndry_eqn}
\end{equation}
which requires that $\tilde B_\kappa(\alpha,z; \bar\alpha,\bar z)^2
-4 \tilde A(\alpha,z; \bar z) \tilde C_\kappa(\bar\alpha,\bar z) \ge 0$.
Thus the step function in $k^{[\ell]}_n(\bar\alpha,\bar z;\alpha,z)$ becomes
\begin{eqnarray}
	 & & \theta\left(
			\bar\alpha-R(\bar z, G(z;y))
			\frac{A(\alpha,z)y^2+B(\alpha,z)+C}{c y+\Delta}
		\right)
	\nonumber \\
	 &  & = \sum_{\kappa=\pm 1}\,
		\theta(\tilde B_\kappa(\alpha,z; \bar\alpha,\bar z)^2
		-4\tilde A(\alpha,z;\bar z)\tilde C_\kappa(\bar\alpha,\bar z))
	 \, \theta(y-y^\kappa_1) \theta(y^\kappa_2-y)
	 \nonumber\\
	 &  & \qquad\qquad\times \theta\left(\kappa
			\left[ (c\bar z -f + b/2)(y+a)
			- b/ 2 \left(\bar z\, b/2-(e-a z) \right)
			\right]	\right)\;,
	\label{step_y}
\end{eqnarray}
where $y^\kappa_i$ with $i=1,2$ are two roots of the
Eq.~(\ref{bndry_eqn})
\begin{equation}
	\left. \matrix{{y^\kappa_1}\cr
			{y^\kappa_2}\cr
	} \right\}
	=\frac{ -\tilde B_\kappa(\alpha,z; \bar\alpha,\bar z)
		\mp \sqrt{\tilde B_\kappa(\alpha,z; \bar\alpha,\bar z)^2
				-4 \tilde A(\alpha,z; \bar z)
				\tilde C_\kappa(\bar\alpha,\bar z) }}
			{2\tilde A(\alpha,z; \bar z)}\;.
	\label{bndry_y}
\end{equation}
The $y$-integration for the functions  $h^{[\ell]}_n(\alpha,z)$,
$k^{[\ell]}_n(\bar\alpha,\bar z;\alpha,z)$ and
$g^{[\ell]}_n(\bar\alpha,\bar z;\alpha,z)$ are
discussed in Appendix \ref{apdx_d}.

Finally we list some useful linear combinations of the
functions $\tilde A(\alpha,z)$ and $B(\alpha,z)$;\\
\begin{eqnarray}
	 & D(z) & \equiv a^2 A(\alpha,z) -a B(\alpha,z) +C \nonumber\\
	 &  &= {b^2\over 4}\left(m^2-{P^2\over 4}\right)
	         +(e-a z)^2 {P^2\over 4},
	\label{d} \\
	 & \tilde D_\kappa(\bar\alpha,\bar z,z)
		& \equiv a^2 \tilde A(\alpha,z; \bar z)
	                       - a \tilde B_\kappa(\alpha,z; \bar\alpha,\bar z)
	                       + \tilde C_\kappa(\bar\alpha,\bar z) \nonumber\\
	 & & = (1-\bar z^2)	D(z)
		+ \left( a \zeta_\kappa(\bar z,z)
				 - \eta_\kappa(\bar	z) \right) \bar\alpha
			\nonumber\\
	 & & = (1-\bar z^2)	D(z)
	     + {b\over 2}\left[{b\over 2}(1+\kappa\bar z)
	                       -(e-a z)(\bar z +\kappa) \right]\bar\alpha\;.
	\label{tilde_d}
\end{eqnarray}

\section{Finite Part of Singular Integrals}\label{apdx_d}

In this Appendix we review the Hadamard finite part prescription
and present the derivation of Eqs.~(\ref{g_krnl}) and (\ref{int_func0}).
We use the following definition
of the Hadamard finite part $\text{Pf}\,\cdot 1/x^n$
together with the $n$-th derivative of the $\delta$-function
\begin{equation}
	\lim_{\epsilon\rightarrow 0}\frac{1}{(x\pm i\epsilon)^n}
	=\text{Pf}\,\cdot 1/x^n
		\mp i\pi
		\frac{(-1)^{n-1}}{(n-1)!}\delta^{(n-1)}(x),
	\label{def_Pf}
\end{equation}
for $n = 1,2,\dots$.
The following formulae are useful;
\begin{eqnarray}
	& & \frac{d}{dx}\text{Pf}\,\cdot 1/x^n =
		(-n)\text{Pf}\,\cdot 1/x^{n+1}\\
	& & x \delta^{\prime}(x)=-\delta(x),\\
	& & x^k \text{Pf}\,\cdot 1/x^n = \text{Pf}\,\cdot 1/x^{n-k},
\end{eqnarray}
for a positive integer $k$.  These are a generalization of original
definition of the Hadamard finite part
\begin{equation}
	\int dx\,\text{Pf}\,\cdot x^\lambda \theta(x) f(x)
	\equiv \lim_{\epsilon\rightarrow 0} \,
	\left[\int_{\epsilon}^\infty x^\lambda f(x)
		+ \sum_{j=0}^k \frac{f^(j)(x)}{j!}
				\frac{\epsilon^{\lambda+j+1}}{\lambda+j+1}
	\right],
	\label{Hadamard's_orgn}
\end{equation}
where $\text{Re}\,\lambda+k+2 > 0$ and $\lambda$ is not a negative integer.
We also apply this prescription for the function
$g^{[\ell]}_n(\bar\alpha,\bar z;\alpha,z)$, as necessary.

Having understood the $\delta$-function as the limit of Eq.~(\ref{def_Pf}),
we shall now derive the Eq.~(\ref{g_krnl}) from Eq.~(\ref{g_func}).
Recalling the Eq.~(\ref{arg_step}), the delta function can be written as
\begin{eqnarray}
	& &\delta\left(\bar\alpha-R(\bar z, G(z;y))
			\frac{A(\alpha,z)y^2+B(\alpha,z)y+C}{c y+\Delta}
		\right)\nonumber\\
	& = &\frac{\zeta_\kappa(\bar z,z)y+\eta_\kappa(\bar z)}{\pi }
	\text{Im}\frac{1}{\tilde A(\alpha,z; \bar z)y^2
			+\tilde B_\kappa(\alpha,z; \bar\alpha,\bar z) y
			+\tilde C_\kappa(\bar\alpha,\bar z)
			-i \epsilon},
\end{eqnarray}
where we have used the fact
$\zeta_\kappa(\bar z,z)y+\eta_\kappa(\bar z)\ge 0$.
Thus the function $g^{[\ell]}_n(\bar\alpha,\bar z;\alpha,z)$ is written as
the $\epsilon\rightarrow 0$ limit
\begin{eqnarray}
	& &g^{[\ell]}_n(\bar\alpha,\bar z;\alpha,z)
	= \frac{1}{(4\pi)^2}
	\text{Pf}\cdot\frac{1}{\bar\alpha}\left(-{b\over 2}\right)^l
	\frac{(1-\bar z^2)^{n+1}}{2}	\nonumber\\
	& &
	\quad\times\text{Re}\,
	\sum\limits_{\kappa=\pm 1}\,\lim_{\epsilon\rightarrow 0}
		\frac{
			\theta(\tilde B_\kappa(\alpha,z; \bar\alpha,\bar z)^2
			 -4 \tilde A(\alpha,z; \bar z)
			 \tilde C_\kappa(\bar\alpha,\bar z))
		}{\left(
			 \tilde B_\kappa(\alpha,z; \bar\alpha,\bar z)^2
			 -4 \tilde A(\alpha,z; \bar z)
			 (\tilde C_\kappa(\bar\alpha,\bar z) -i\epsilon)
		\right)^{1/2}}
		\sum\limits_{i=1,2}\,
		\frac{
			{\bar y^{\kappa\,n+1}_i}\,(a+\bar y^\kappa_i)^{n-1-l}
		}{
			\left(\zeta_\kappa(\bar z,z) \bar y^\kappa_i
				 + \eta_\kappa(\bar z) \right)^n
		}
\nonumber\\
& & \qquad\quad\times
		\theta(\text{Re}\bar y^\kappa_i)
		\theta\left(\kappa \text{Re}
			\left[ \left(c\bar z -f + {b\over 2}\right)
				(\bar y^\kappa_i+a)
			- {b\over 2}\left(\bar z\, {b\over 2}-(e-a z) \right)
			\right]
		\right),
	\label{exp_reg_g_func}
\end{eqnarray}
where $\bar y^\kappa_i$ are root of the equation;
\begin{equation}
	\tilde A(\alpha,z; \bar z)y^2
			+\tilde B_\kappa(\alpha,z; \bar\alpha,\bar z) y
			+\tilde C_\kappa(\bar\alpha,\bar z)
			-i \epsilon = 0.
\end{equation}
It is now easy to rewrite this expression of the form Eq.~(\ref{g_krnl}).

Now, let us turn to the evaluation of the integral Eq.~(\ref{func0});
\begin{equation}
	\tilde I(y_{\rm min},y_{\rm max})\equiv
	\int_{y_{\rm min}}^{y_{\rm max}}dy\,\frac{y}{y+\tilde a}
		\frac{1}{A y^2 + B y +C}.
\end{equation}
Changing the integration variable such that
\begin{eqnarray}
	& & Y=\left(\frac{\tilde a}{y+\tilde a}\right)^2(Ay^2+By+C),\\
	& & X=\frac{\sqrt{B^2-4AC}\, y}{2C+By},
\end{eqnarray}
the integral for $B^2-4AC \ge 0$ can be written as follows;
\begin{eqnarray}
	& \tilde I(y_{\rm min},y_{\rm max}) =
	{1 \over \tilde a^2 A -\tilde a B + C}
	&\left\{ \frac{\tilde a}{4}
		\int_{Y_{\rm min}}^{Y_{\rm max}}dY\,\text{Pf}\cdot\,
                {1\over Y}\right.
	\nonumber\\
	& &\quad  \left.
	+\frac{2C-\tilde a B}{ 4\sqrt{B^2-4AC} }
		\int_{X_{\rm min}}^{X_{\rm max}} dX\,
			\left( \text{Pf}\cdot\,{1\over 1+X}
				-\text{Pf} \cdot\,{1\over 1-X} \right)
	\right\},
\end{eqnarray}
where $Y_{\rm min}, X_{\rm min}, \dots$ are the values of $Y$ and $X$
at $y=y_{\rm min},y_{\rm max}$.  It is easy to
derive the Eq.~(\ref{int_func0})
by performing the integral over $X$ and $Y$ with
the definition of the Hadamard finite part (\ref{def_Pf}).



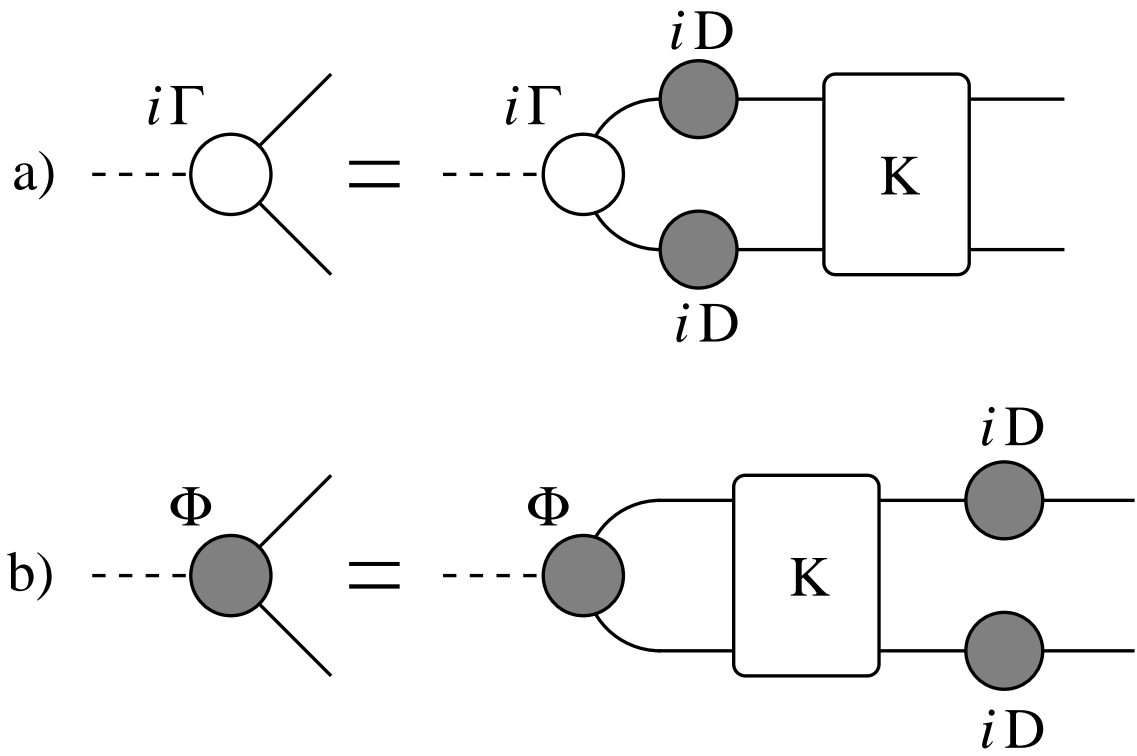
\begin{figure}[tb]
  \centering
  \parbox{130mm}{\caption{The Bethe-Salpeter (BS) equation for two
    scalar constituents in terms of the bound state proper vertex
    (a) and in terms of the BS amplitude (b).
  \label{bse_fig} }}
  \vspace{1.0cm}
\end{figure}

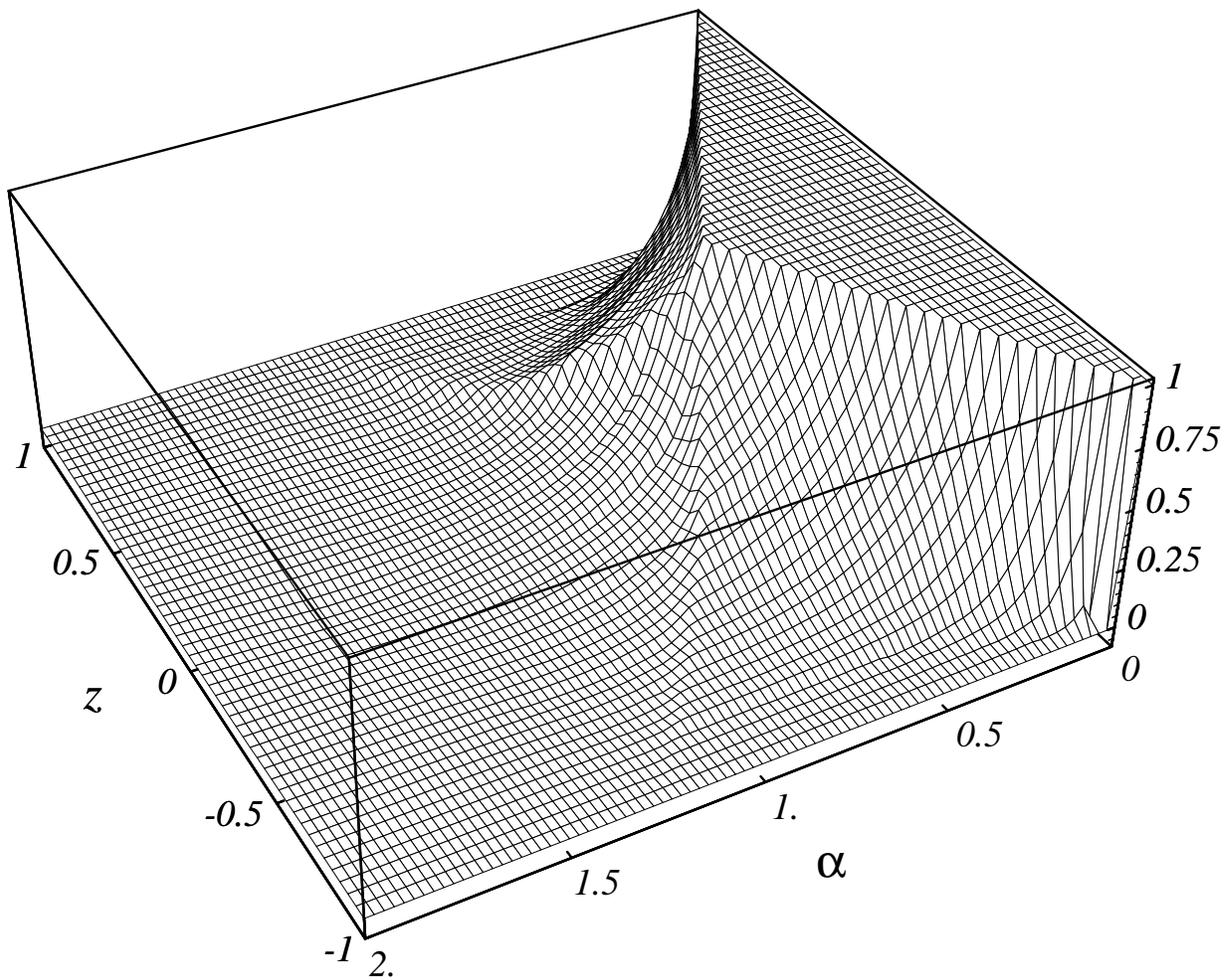
\begin{figure}[tb]
  \centering
  \parbox{130mm}{\caption{
    The s-wave ($\ell=0$) Bethe-Salpeter (BS) amplitude
    weight function for $n=1$, $\phi^{[\ell=0]}_1(\alpha,z)$,
    for the pure ladder kernel
    function described in the text.
  \label{fig_pure} }}
  \vspace{1.0cm}
\end{figure}

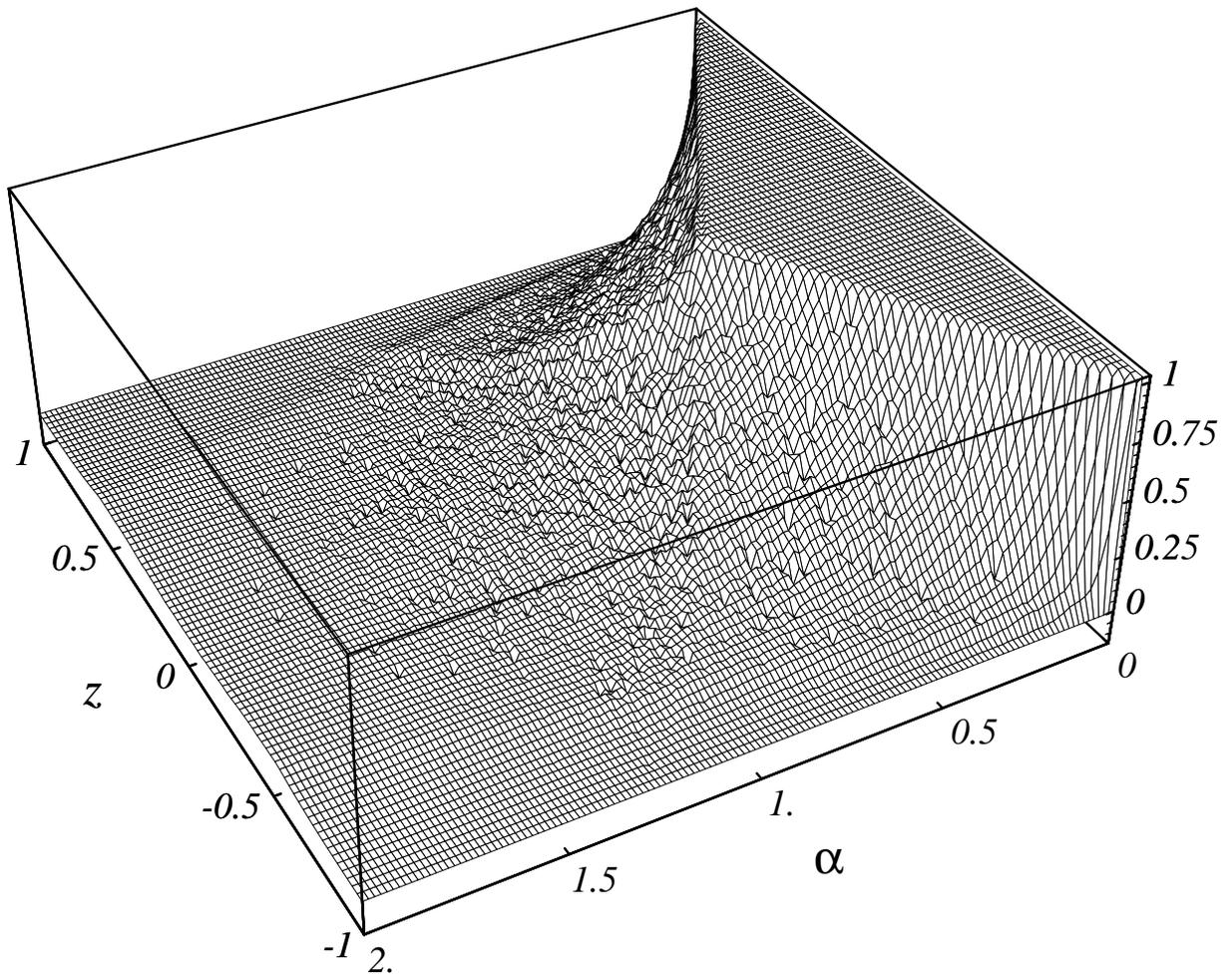
\begin{figure}[tb]
  \centering
  \parbox{130mm}{\caption{
    The s-wave ($\ell=0$) Bethe-Salpeter (BS) amplitude
    weight function for $n=1$, $\phi^{[\ell=0]}_1(\alpha,z)$,
    for the example general kernel function
    described in the text.
  \label{fig_gnrl} }}
  \vspace{1.0cm}
\end{figure}

\end{document}